\documentclass[a4paper,11pt,reqno]{amsart}

\usepackage[margin=35mm]{geometry}
\usepackage[T1]{fontenc}
\usepackage[english]{babel}
\usepackage[utf8]{inputenc}
\usepackage{amsmath,amssymb,amsthm,amsfonts,amstext,amsopn,amsxtra,dsfont,mathrsfs,esint,enumitem,bookmark,mathtools}
\usepackage{bbm}
\usepackage[capitalise]{cleveref}
\hypersetup{linktocpage=true}
\usepackage[headings]{fullpage}
\usepackage{microtype}

\usepackage{comment}

\newcommand\numberthis{\addtocounter{equation}{1}\tag{\theequation}}

\theoremstyle{plain}
\newtheorem{theorem}{Theorem}
\newtheorem{lemma}[theorem]{Lemma}
\newtheorem{proposition}[theorem]{Proposition}

\theoremstyle{definition}

\newcommand\xqed[1]{%
	\leavevmode\unskip\penalty9999 \hbox{}\nobreak\hfill\quad\hbox{#1}%
}
\newcommand\remarkend{\xqed{$\triangle$}}

\makeatletter
\def\@endtheorem{\remarkend\endtrivlist\@endpefalse }
\makeatother
\theoremstyle{remark}
\newtheorem{remark}[theorem]{Remark}
\makeatletter
\def\@endtheorem{\endtrivlist\@endpefalse }
\makeatother

\crefname{theorem}{Theorem}{Theorems}
\crefname{lemma}{Lemma}{Lemmas}
\crefname{proposition}{Proposition}{Propositions}
\crefname{corollary}{Corollary}{Corollaries}
\crefname{definition}{Definition}{Definitions}
\crefname{assumption}{Assumption}{Assumptions}
\crefname{remark}{Remark}{Remarks}

\crefname{subsection}{subsection}{subsections}
\crefname{subsubsection}{subsection}{subsections}

\renewcommand{\d}[1]{\ensuremath{\operatorname{d}\!{#1}}}

\newcommand{\Tr}{\operatorname{Tr}}

\newcommand{\e}{\operatorname{e}}
\newcommand{\op}{\operatorname{op}}
\newcommand{\A}{\mathbf{A}}

\renewcommand{\i}{\mathbf{i}}
\newcommand{\p}{\mathbf{p}}

\newcommand{\R}{\mathbb{R}}
\newcommand{\af}{\textmd{\normalfont af}}
\newcommand{\imax}{i_{\textmd{\normalfont max}}}
\newcommand{\jmax}{j_{\textmd{\normalfont max}}}
\newcommand{\bbP}{\mathbb{P}}
\newcommand{\bbQ}{\mathbb{Q}}
\newcommand{\cE}{\mathcal{E}}
\newcommand{\cS}{\mathcal{S}}
\newcommand{\fH}{\mathfrak{H}}

\newcommand{\sym}{\textmd{\normalfont s}}
\newcommand{\osym}{\otimes_\sym}

\newcommand{\mfH}{\mathfrak{H}}

\newcommand{\bbN}{\mathbb{N}}

\newcommand\mydots{\ifmmode\mathellipsis\else.\kern-0.08em.\kern-0.08em.\fi}

\allowdisplaybreaks

\DeclareUnicodeCharacter{2217}{*}

\usepackage{csquotes}
\usepackage[backend=bibtex,sorting=nty,citestyle=numeric-comp,bibstyle=ieee,maxbibnames=99,doi=false,isbn=false,url=false,eprint=false,dashed=true]{biblatex}
\AtEveryBibitem{\clearfield{month}}
\AtEveryBibitem{\clearfield{day}}
\usepackage{graphicx}
\usepackage{subcaption}
\usepackage[table,xcdraw]{xcolor}

\usepackage{tikz}
\usetikzlibrary{arrows}
\usepackage{tkz-tab}

\usepackage{standalone}

\usepackage{hyperref}
\hypersetup{bookmarksdepth=3}
\DeclareRobustCommand{\SkipTocEntry}[9]{}

\title[Average-field approximation almost-bosonic anyons]{Average-field approximation for dilute almost-bosonic anyons}
\author[F. L. A. Visconti]{François L. A. Visconti}

\address{Institute of Science and Technology Austria, Am Campus 1, 3400 Klosterneuburg, Austria}
\email{francois.visconti@ist.ac.at}

\begin{document}
	\maketitle

	\begin{abstract}
		We study the ground state of a system of $N$ two-dimensional trapped almost-bosonic anyons. This setup can equivalently be viewed as bosons interacting through long-range magnetic potentials generated by magnetic charges carried by each particle. These magnetic charges are assumed to be  smeared over discs of radius $R$---a model known as \textit{extended anyons}. To recover the point-like anyons perspective, we consider the joint limit $R \rightarrow 0$ as $N \rightarrow \infty$. We rigorously justify the \textit{average-field approximation} for any radius $R$ that decays polynomially, and even for certain exponentially decaying $R$. The average-field approximation asserts that the particles behave like independent, identical bosons interacting through a self-consistent magnetic field. Our result significantly improves upon the best-known estimates from \cite{Girardot2020averageFA,Lundholm2015AverageFA}, and it in particular covers radii that are much smaller than the mean interparticle distance. The proof strategy builds on recent work on two-dimensional attractive Bose gases by Junge and the author \cite{Junge2025DerivationHT2D}.
	\end{abstract}
		
	\tableofcontents
	
	\section{Introduction}
	
	A system of $N$ free two-dimensional anyons with statistics parameter $\alpha$ can \textit{formally} be described by an $N$-body wavefunction of the form
	\begin{equation*}
		\tilde{\Psi}(x_1,\dots,x_N) = \prod_{1\leq j< k\leq N}e^{\mathbf{i}\alpha\phi_{jk}}\Psi(x_1,\dots,x_N) \quad \textmd{with} \quad \phi_{jk} = \arg\dfrac{x_j - x_k}{\vert x_j - x_k\vert},
	\end{equation*}
	where $\Psi$ is a bosonic wavefunction, which is to say symmetric under particle exchange. We refer to \cite{Khare98,Lundholm2024Properties2DA,Myrheim02,Ouvry09,Wilczek90} for physical references. Heuristically, the wavefunction $\tilde{\Psi}$ satisfies
	\begin{equation*}
		\tilde{\Psi}(x_1,\dots,x_j,\dots,x_k,\dots,x_N) = e^{\mathbf{i}\alpha\pi}\tilde{\Psi}(x_1,\dots,x_k,\dots,x_j,\dots,x_N),
	\end{equation*}
	meaning that its behaviour under particle exchange interpolates between bosons ($\alpha = 0$) and fermions ($\alpha = 1$). However, this implies that the wavefunction is in general not single-valued, making this description difficult to use in practice. A way to overcome this issue is to realise that the action of the free Hamiltonian
	\begin{equation*}
		\sum_{j = 1}^N-\Delta_j + V(x_j)
	\end{equation*}
	on the anyonic wavefunction $\tilde{\Psi}$ is \textit{formally} equivalent to the action of an effective, $\alpha$-dependent Hamiltonian on the bosonic wavefunction $\Psi$ \cite{Lundholm2013HardyLT,Lundholm2024Properties2DA,Myrheim2002Anyons}. This is called the magnetic gauge picture of the anyonic problem and consists in working with the Hamiltonian
	\begin{equation}
		\label{eq:hamiltonian_point_like}
		H_N = \sum_{j = 1}^N\left(\mathbf{p}_j +  \alpha\A_j\right)^2 + V(x_j)
	\end{equation}
	acting on
	\begin{equation*}
		\mathfrak{H}^N = \bigotimes_{\textmd{sym}}^N\mathfrak{H},
	\end{equation*}
	the symmetric tensor product of $N$ copies of the one-body Hilbert space $\mathfrak{H} \coloneqq L^2(\mathbb{R}^2)$. Here, we introduced the momentum operator
	\begin{equation*}
		\p_j = -\mathbf{i}\nabla_j,
	\end{equation*}
	and the statistical gauge vector potential
	\begin{equation}
		\label{eq:potential_singular}
		\mathbf{A}_j = \sum_{\substack{k = 1\\ k\neq j}}^N\dfrac{\left(x_j - x_k\right)^\perp}{\vert x_j - x_k\vert^2}
	\end{equation}
	felt by the $j$-th particle due to the influence of all other particles (denoting $(x,y)^\perp = (-y,x)$). We assume that the external trapping potential $V:\mathbb{R}^2 \rightarrow \mathbb{R}$ satisfies\footnote{In all this work, we denote by $c$ or $C$ generic nonnegative constants that might differ from line to line.}
	\begin{equation}
		\label{eq:trapping_potential_assumption}
		V(x) \geq c^{-1}|x|^s - C
	\end{equation}
	for some constant $s > 0$, and that its gradient is such that
	\begin{equation}
		\label{eq:trapping_potential_regularity_assumption}
		\vert\nabla V\vert \leq C + cV^{1/2} + \xi V^{3/2},
	\end{equation}
	for some constant $0\leq \xi < \sqrt{2}$. The last condition ensures that \cite[Theorem 3.23]{Lewin24} holds.
		
	We consider $\alpha$ of the form
	\begin{equation}
		\label{eq:statistics_parameter}
		\alpha = \frac{\beta}{N - 1},
	\end{equation}
	for some given fixed constant $\beta$. This places us in a so-called \textit{almost-bosonic} limit, because $\alpha \rightarrow 0$ as $N \rightarrow \infty,$ and it ensures that the anyon statistics has an effect at the leading order \cite{Ataei2025microscopicDA,Correggi2017LocalDA,Correggi2019VorexPAB,Girardot2020averageFA,Girardot2025derivationCSS,Lundholm2015AverageFA}.
	
	\subsection{Extended anyons}
	
	The Hamiltonian \eqref{eq:hamiltonian_point_like} is actually too singular to be considered as acting on product states $u^{\otimes N}$, regardless of the regularity of the one-particle wavefunction $u$. To resolve this, we introduce a length scale $R$ over which the magnetic charges are smeared; this is called an \textit{extended anyons} model. For a detailed discussion, we refer to \cite{Larson2017ExclusionBEA,Lundholm2015AverageFA,Lundholm2017ManyATS} and references therein. We let $R \rightarrow 0$ when $N \rightarrow \infty$ to recover the point-like anyons perspective.
	
	Let $w_R$ be the two-dimensional Coulomb potential generated by a unit charge smeared over a disc of radius $R$:
	\begin{equation}
		\label{eq:smeared_coulomb_potential}
		w_R = \log\vert\cdot\vert *\dfrac{\mathds{1}_{B(0,R)}}{\pi R^2}.
	\end{equation}
	Define also $w_0 = \log \vert\cdot\vert$. Considering that
	\begin{equation*}
		\nabla^\perp w_0(x) = \dfrac{x^\perp}{\vert x\vert^2},
	\end{equation*}
	it is natural to introduce the regularised gauge vector potential $\A_j^R$ given by
	\begin{equation}
		\label{eq:potential_extended}
		\mathbf{A}_j^R = \sum_{\substack{k = 1\\ k\neq j}}^N\nabla^\perp w_R(x_j - x_k).
	\end{equation}
	This is associated to the regularised Hamiltonian
	\begin{equation}
		\label{eq:hamiltonian_regularised}
		H_N^R = \sum_{j = 1}^N\left(\mathbf{p}_j + \alpha\mathbf{A}_j^R\right)^2 + V(x_j).
	\end{equation}
	For fixed $R > 0$, this operator is essentially self-adjoint and bounded from below (see \cite{Avron1978SchroedingerOMF},\cite[Theorem X.17]{Reed1978MethodsMMP2}) and can be expanded as
	\begin{multline*}
		H_N^R = \sum_{j = 1}^N-\Delta_j + V(x_j) + 2\alpha\sum_{\substack{1\leq j,k\leq N\\ j\neq k}}\mathbf{p}_j\cdot \nabla^\perp w_R(x_j - x_k)\\
		+ \alpha^2\sum_{\substack{1\leq j,k,\ell\leq N\\ j\neq k\neq \ell\neq j}}\nabla^\perp w_R(x_j - x_k)\cdot\nabla^\perp w_R(x_j - x_\ell) + \alpha^2\sum_{\substack{1\leq j,k\leq N\\ j\neq k}}\left\vert\nabla w_R(x_j - x_k)\right\vert^2. \numberthis \label{eq:hamiltonian_regularised_rewritten}
	\end{multline*}
	Here we used that $\p$ and $\nabla^\perp w_R$ commute. Though this operator does not have a unique limit as $R \rightarrow 0$ and the singular Hamiltonian \eqref{eq:hamiltonian_point_like} is not essentially self-adjoint \cite{Adami1998AharonovBE,Bourdeau1992WhenIPC,Corregi2018Hamiltonians2AS,Dabrowski1998AharonovBohm,Lundholm2014LocalELT}, we obtain a unique nonlinear model when taking the joint limit $R\rightarrow 0$ as $N\rightarrow\infty$ as long as $R \ll e^{-N}$. We refer to \cite{Ataei2025microscopicDA} for a discussion on more singular regimes.
	
	\subsection{Average-field approximation}
	
	The average-field approximation is obtained by substituting the potentials \eqref{eq:potential_singular} and \eqref{eq:potential_extended} with the average-field potentials $\A[\rho]$ and $\A^R[\rho]$ defined by
	\begin{equation*}
		\mathbf{A}[\rho] = \nabla^\perp w_0*\rho \quad \textmd{and} \quad \mathbf{A}^R[\rho] = \nabla^\perp w_R*\rho,
	\end{equation*}
	where $\rho$ is the \textit{normalised} 1-body density of a given wavefunction $\Psi$:
	\begin{equation*}
		\rho(x) = \int_{\mathbb{R}^{2(N - 1)}}\d{}x_2\dots\d{}x_N\vert\Psi(x,x_2,\dots,x_N)\vert^2.
	\end{equation*}
	Then, the regularised Hamiltonian \eqref{eq:hamiltonian_regularised} is approximated by the $N$-body average-field Hamiltonian
	\begin{equation*}
		H_N^{\textmd{af}}[\rho] = \sum_{j = 1}^N\left(\mathbf{p}_j + \alpha (N - 1)\mathbf{A}^R[\rho]\right)^2 + V(x_j).
	\end{equation*}
	At fixed density $\rho$, this is a noninteracting magnetic Hamiltonian acting on $\mathfrak{H}^N$, so it is natural to use the ansatz $u^{\otimes N}$, which has energy per particle
	\begin{equation*}
		N^{-1}\left\langle u^{\otimes N},H_N^{\textmd{af}}[\rho]u^{\otimes N}\right\rangle = \left\langle u, \left[(\mathbf{p} + \alpha (N - 1)\mathbf{A}^R[\rho])^2 + V\right] u\right\rangle.
	\end{equation*}
	For consistency, we set $\rho = \vert u\vert^2$, meaning that the density is that of the state $u^{\otimes N}$. This leads to the \textit{nonlinear average-field energy functional} $\cE_R^\af$ given by
	\begin{equation}
		\label{eq:average_field_energy_functional}
		\mathcal{E}_R^{\textmd{af}}[u] = \int_{\mathbb{R}^2}\left(\left\vert(\p + \mathbf{i}\beta \mathbf{A}^R[\vert u\vert ^2])u\right\vert^2 + V\vert u\vert^2\right),
	\end{equation}
	where $\alpha (N - 1)$ was replaced by $\beta$ in accordance with \eqref{eq:statistics_parameter}. Furthermore, in the limit $R \rightarrow 0$ this energy functional converges to the \textit{singular average-field energy functional} $\cE^\af$ given by
	\begin{equation}
		\label{eq:average_field_energy_functional_singular}
		\mathcal{E}^{\textmd{af}}[u] = \int_{\mathbb{R}^2}\left(\left\vert(\p + \mathbf{i}\beta \mathbf{A}[\vert u\vert ^2])u\right\vert^2 + V\vert u\vert^2\right)
	\end{equation}
	(see \cite[Proposition A.5]{Girardot2020averageFA},\cite[Proposition A.6]{Lundholm2015AverageFA}).
	
	\subsection{The mean-field point of view}
	
	In principle, the average-field approximation does not require the Hamiltonian $H_N^R$ to exhibit Bose--Einstein condensation. In fact, it has been widely used in the physics literature to perturb around fermions ($\alpha = 1$) \cite{Fetter1989RandomPA,GirardotRou21,GirardotRou23,Iengo1992AnyonQM,LevittLunRou26,Trugenberger1992AnyonFB,Trugenberger1992groundSCE,Westerberg1993meanFAA}. However, when taking $\beta = \alpha (N - 1)$ fixed, which is natural for the study of $\cE_R^\af$, the limit $N \rightarrow \infty$ can be seen as a mean-field-like limit of many-body bosonic systems. Indeed, the two-body terms in \eqref{eq:hamiltonian_regularised_rewritten} come with prefactors $1/N$ and $1/N^2$, while the three-body term with a prefactor $1/N^2$. See \cite[Section 1.4]{Rougerie20} for a related discussion.
	
	Furthermore, the functionals $\cE_R^\af$ and $\cE^\af$ defined in \eqref{eq:average_field_energy_functional} and \eqref{eq:average_field_energy_functional_singular} may also be derived from the mean-field picture of $H_N^R$. To see this, we write, for some ground state $\Psi$ of $H_N^R$,
	\begin{multline}
		\label{eq:energy_rewritten_trace}
		\left\langle\Psi,H_N^R\Psi\right\rangle/N = \Tr\big((-\Delta + V)\Gamma^{(1)}\big) + 2\beta\Tr\big(\p_1\cdot\nabla^\perp w_R(x_1 - x_2)\Gamma^{(2)}\big)\\
		+ \beta^2\dfrac{N - 2}{N - 1}\Tr\big(\nabla^\perp w_R(x_1 - x_2)\cdot\nabla^\perp w_R(x_1 - x_3)\Gamma^{(3)}\big)\\
		+ \beta^2\dfrac{1}{N - 1}\Tr\big(\left\vert\nabla w_R(x_1 - x_2)\right\vert^2\Gamma^{(2)}\big),
	\end{multline}
	where $\Gamma^{(k)}$ denotes the $k$-body reduced density matrix of $\Psi$. Since we are considering a mean-field bosonic Hamiltonian (for fixed $R$ at least), it is natural to use the ansatz
	\begin{equation*}
		\Psi \approx u^{\otimes N} \quad \textmd{and} \quad \Gamma^{(k)} \approx \vert u^{\otimes k}\rangle\langle u^{\otimes k}\vert,
	\end{equation*}
	for which
	\begin{equation}
		\label{eq:two_body_term_pure_product_rewritten}
		\Tr\big(\mathbf{p}_1\cdot\nabla^\perp w_R(x_1 - x_2)\vert u^{\otimes 2}\rangle\langle u^{\otimes 2}\vert\big) = \int_{\mathbb{R}^2}\mathbf{A}^R[\vert u\vert^2]\cdot u\overline{\p u}
	\end{equation}
	and
	\begin{equation}
		\label{eq:three_body_term_pure_product_rewritten}
		\Tr\big(\nabla^\perp w_R(x_1 - x_2)\cdot\nabla^\perp w_R(x_1 - x_3)\vert u^{\otimes 3}\rangle\langle u^{\otimes 3}\vert\big) = \int_{\mathbb{R}^2}\vert u\vert^2\left\vert\mathbf{A}^R[\vert u\vert^2]\right\vert^2.
	\end{equation}
	Inserting these identities into \eqref{eq:energy_rewritten_trace} and neglecting the last term because it is heuristically of order $\mathcal{O} (N^{-1})$, we obtain
	\begin{equation*}
		N^{-1}\langle\Psi,H_N^R\Psi\rangle \approx \mathcal{E}_R^{\textmd{af}}[u].
	\end{equation*}
	Taking the joint limit $R\rightarrow 0$ as $N\rightarrow \infty$ again leads to the singular functional $\cE^\af$.
	
	The convergence of the ground state energy per particle of $H_N^R$ to that of the functional $\cE^\af$, as well as Bose--Einstein condensation in its minimisers, were first established in \cite{Lundholm2015AverageFA} for anyons with radius $R = N^{-\eta}$, under the condition $0 < \eta < \frac{1}{4}(1 + \frac{1}{s})^{-1}$, where $s$ is the exponent in \eqref{eq:trapping_potential_assumption}. This range was later extended to $0 < \eta < 1/4$ in \cite{Girardot2020averageFA}, where an external magnetic field was considered. In the present paper we extend these results to anyons of radius $N^{-\eta}$ for any $\eta > 0$, and even to anyons of radius $R = e^{-N^\kappa}$ for any $0 < \kappa < 1$. In particular, we cover the case $R \ll N^{-1/2}$, meaning that the radius of the magnetic charges is much smaller than the mean interparticle distance.
	
	For anyons of radius $R = e^{-N^\kappa}$ with $\kappa \geq 1$, our main result is no longer expected to hold. In this regime, the energy of the fully uncorrelated trial state $\Psi = u^{\otimes N}$ diverges due to the singular two-body interaction, regardless of how smooth $u$ is. This indicates that correlations must be taken into account in the trial state, leading to a different energy functional than $\cE^\af$. Recently, an upper bound on the energy---which even covers the singular case $R = 0$---was derived in \cite{Ataei2025microscopicDA} using Jastrow--Dyson-type trial states. Deriving a matching lower bound remains an interesting but likely challenging open problem.
	
	\subsection{Main result}
	
	Define the many-body ground state energy per particle
	\begin{equation}
		\label{eq:ground_state_energy_many_body}
		e_N^R = N^{-1}\inf\left\{\left\langle\Psi,H_N^R\Psi\right\rangle: \Psi\in\mathfrak{H}^{N}, \left\Vert\Psi\right\Vert = 1\right\},
	\end{equation}
	and the ground state energy of the average-field functional \eqref{eq:average_field_energy_functional_singular}
	\begin{equation}
		\label{eq:ground_state_energy_af_singular}
		e^{\textmd{af}} = \inf\left\{\mathcal{E}^{\textmd{af}}[u]: u\in\mathfrak{H}, \Vert u\Vert = 1\right\}.
	\end{equation}
	We prove the convergence of $e_N^R$ to $e^{\textmd{af}}$.  The convergence of the ground states of $H_N$ to condensates in minimisers of $\mathcal{E}^{\textmd{af}}$ follows directly from the convergence of the energies, thanks to arguments from \cite{Lewin2014TheMA} (see also \cite{Girardot2020averageFA,Lundholm2015AverageFA}).
	\begin{theorem}
		\label{th:convergence_gse_average_field}
		Consider $N$ extended anyons of radius $R$ in an external potential $V$ satisfying \eqref{eq:trapping_potential_assumption}. Suppose that $R$ is either of the form 
		\begin{equation}
			\label{eq:smearing_radius_polynomial}
			R = N^{-\eta},
		\end{equation}
		for some $\eta > 0$, or of the form
		\begin{equation}
			\label{eq:smearing_radius_exponential}
			R = e^{-N^\kappa},
		\end{equation}
		for some $0 < \kappa < 1$. Take the statistics parameter to scale as
		\begin{equation}
			\label{eq:statistics_parameter_main_theorem}
			\alpha = \dfrac{\beta}{N - 1}
		\end{equation}
		for some fixed $\beta \in\mathbb{R}$. Define $e_N^R$ and $e^\af$ as in \eqref{eq:ground_state_energy_many_body} and \eqref{eq:ground_state_energy_af_singular}. Then,
		\begin{equation}
			\label{eq:energy_converence_main_result}
			\boxed{\lim_{N\rightarrow \infty}e_N^R = e^{\normalfont\textmd{af}} > -\infty.}
		\end{equation}
		Moreover, for a sequence $\{\Psi_N\}_N$ of ground states of $H_N^R$ with $k$-particle reduced density matrices $\{\Gamma_N^{(k)}\}_N$, there exists a Borel probability measure $\mu$ supported on the set of minimisers of $\mathcal{E}^{\textmd{\normalfont af}}$ such that, along a subsequence,
		\begin{equation}
			\label{eq:state_convergence_main_result}
			\lim_{N\rightarrow\infty}\Tr\Big\vert\Gamma_N^{(k)} - \int\d{}\mu(u)\vert u^{\otimes k}\rangle\langle u^{\otimes k}\vert\Big\vert = 0,. 
		\end{equation}
		for any integer $k$. If $\mathcal{E}^{\textmd{\normalfont af}}$ has a unique minimiser $u_{\textmd{\normalfont af}}$ (up to a phase), then, for the whole sequence,
		\begin{equation*}
			\lim_{N \rightarrow \infty}\Tr\Big\vert\Gamma_N^{(k)} - \vert u_{\textmd{\normalfont af}}^{\otimes k}\rangle\langle u_{\textmd{\normalfont af}}^{\otimes k}\vert\Big\vert = 0,
		\end{equation*}
		for any integer $k$.
	\end{theorem}
	
	\begin{remark}
		Instead of considering free anyons, we can introduce an interaction potential into the Hamiltonian \eqref{eq:hamiltonian_regularised}, as was done in \cite{Nguyen20242Dattractive}. Specifically, we consider the Hamiltonian
		\begin{equation*}
			H_{N,R}^{\textmd{\normalfont int}} = \sum_{j = 1}^N\left(\mathbf{p}_j +  \alpha\A_j^R\right)^2 + V(x_j) + \dfrac{1}{N - 1}\sum_{1\leq i< j\leq N}v_N(x_i - x_j)
		\end{equation*}
		acting on $\mathfrak{H}^N$, where the interaction potential $v_N$ is of the form
		\begin{equation*}
			v_N = N^{2\nu}v(N^\nu\cdot) \quad \textmd{or} \quad v_N = e^{2N^\nu}v(e^{N^\nu}\cdot),
		\end{equation*}
		for some $\nu > 0$ (with $\nu < 1$ in the exponential case). Using the uncorrelated trial state $\Psi = u^{\otimes N}$ leads to the singular functional
		\begin{equation}
			\label{eq:average_field_energy_functional_singular_interaction}
			\cE^{\textmd{\normalfont int}}[u] = \int_{\R^2}\left(\left\vert(\p + \i\beta \A[\vert u\vert^2])u\right\vert^2 + V\vert u\vert^2 + \dfrac{a}{2}\vert u\vert^4\right), 
		\end{equation}
		with $a = \int_{\R^2}v$. To ensure that the functional \eqref{eq:average_field_energy_functional_singular_interaction} is stable, one needs to assume that
		\begin{equation*}
			a > -a^*(\beta),
		\end{equation*}
		where $a^*(\beta)$ is the optimal constant in the magnetic Gagliardo--Nirenberg inequality:
		\begin{equation*}
			\dfrac{a^*(\beta)}{2}\int_{\R^2}\vert u\vert^4 \leq \int_{\R^2}\left\vert(\p + \beta\A[\vert u\vert^2])u\right\vert^2,
		\end{equation*}
		valid for all $u\in H^1(\R^2)$ with $\Vert u\Vert = 1$
		(see \cite{Ataei2025GeneralizedLEM,Nguyen20242Dattractive} for more details). Assuming $v\in L^1(\R^2)\cap L^{1 + \varepsilon}(\R^2)$ for some $\varepsilon > 0$, and taking $R$ as in Theorem~\ref{th:convergence_gse_average_field}, a straightforward combination of the proof of Theorem~\ref{th:convergence_gse_average_field} and the proof of \cite[Theorem 2]{Junge2025DerivationHT2D} yields
		\begin{equation}
			\label{eq:energy_converence_interaction}
			\lim_{N \rightarrow \infty}e_{N,R}^{\textmd{\normalfont int}} = e^{\textmd{\normalfont int}},
		\end{equation}
		where $e_{N,R}^{\textmd{\normalfont int}}$ is the ground state energy per particle of $H_{N,R}^{\textmd{\normalfont int}}$ and $e^{\textmd{\normalfont int}}$ is the ground state energy of $\cE^{\textmd{\normalfont int}}$. Bose--Einstein condensation (BEC) in minimisers of $\cE^{\textmd{\normalfont int}}$ follows from the energy convergence. In \cite{Nguyen20242Dattractive}, the convergence \eqref{eq:energy_converence_interaction} (and consequently BEC) was shown for potentials of the form $v_N = N^{2\nu}v(N^\nu\cdot)$ and for $R = N^{-\eta}$, under the much more restrictive conditions $\nu < \frac{1}{6}(1 + \frac{1}{s})^{-1}$ and $\eta < \frac{1}{4}(1 + \frac{1}{s})^{-1}$, where $s$ is the exponent in \eqref{eq:trapping_potential_assumption}.
	\end{remark}
	
	\bigskip
	\noindent
	\textbf{Organisation of the paper.} In Section~\ref{sec:notation_preliminaries_proof_strategy}, we fix some notation, provide preliminary results, and explain the proof strategy of Theorem~\ref{th:convergence_gse_average_field}. In Section~\ref{section:proof_main_theorem}, we prove Theorem~\ref{th:convergence_gse_average_field} assuming the validity of two key estimates: Lemmas~\ref{lemma:high_momenta_estimates}~and~\ref{lemma:low_occupancy_estimates}. Section~\ref{section:proof_main_estimates} is entirely devoted to the proof of Lemmas~\ref{lemma:high_momenta_estimates}~and~\ref{lemma:low_occupancy_estimates}.
	
	\bigskip
	\noindent
	\textbf{Acknowledgements.} We express our sincere gratitude to Nicolas Rougerie for the introduction to the problem, for his hospitality, and for valuable discussions. We thank Robert Seiringer for discussions surrounding Proposition \ref{prop:plane_wave_estimate}. We also thank Douglas Lundholm, Théotime Girardot and the anonymous referee for their valuable feedback. Partial support by the Deutsche Forschungsgemeinschaft (DFG, German Research Foundation) through the TRR 352 Project ID 470903074 and by the European Research Council through the ERC CoG RAMBAS Project Nr. 101044249 is acknowledged.
	
	\section{Notation, preliminaries and proof strategy}
	
	\label{sec:notation_preliminaries_proof_strategy}
	
	\subsection{Notation}
	
	Define the one-body operator
	\begin{equation}
		\label{eq:one_body_operator_def}
		h = -\Delta + V.
	\end{equation}
	Define the three-body Hamiltonian
	\begin{equation}
		\label{eq:three_body_hamiltonian}
		\tilde{H}_3^R = T + \beta W_2 + \beta^2W_3,
	\end{equation}
	with the operators $T,W_2$ and $W_3$ given by
	\begin{equation}
		\label{eq:effective_hamiltonian_three_terms_def}
		T = \dfrac{1}{3}(h_1 + h_2 + h_3), \quad W_2 = \dfrac{1}{3}\sum_{\substack{1\leq j,k\leq 3\\ j\neq k}}W_2(j,k) \quad \textmd{and} \quad W_3 = \dfrac{1}{6}\sum_{\substack{1 \leq j,k,\ell\leq 3\\ j\neq k\neq \ell\neq j}}W_3(j,k,\ell),
	\end{equation}
	with
	\begin{equation}
		\label{eq:two_and_three_body_term_def}
		W_2(j,k) = \p_j\cdot\nabla^\perp w_R(x_j - x_k) \quad \textmd{and} \quad W_3(j,k,\ell) = \nabla^\perp w_R(x_j - x_k)\cdot\nabla^\perp w_R(x_j - x_\ell).
	\end{equation}
	
	Let $S_N$ be the group of permutations of $\left\{1,\dots,N\right\}$. For $\psi_1 \in \mathfrak{H}^{N_1}$ and $\psi_2 \in \mathfrak{H}^{N_2}$, define the symmetric tensor product $\psi_1\otimes_\textmd{s}\psi_2\in\mathfrak{H}^{N_1 + N_2}$ by
	\begin{equation*}
		\psi_1\otimes_\textmd{s}\psi_2(x_1,\dots,x_{N_1 + N_2}) = \dfrac{1}{\sqrt{N_1!N_2!(N_1 + N_2)!}}\sum_{\sigma\in S_{N_1 + N_2}}
		\begin{multlined}[t]
			\psi_1(x_{\sigma(1)},\dots,x_{\sigma(N_1)})\\
			\times\psi_2(x_{\sigma(N_1 + 1)},\dots,x_{\sigma(N_1 + N_2)}).
		\end{multlined}
	\end{equation*}
	Denote the $i$-fold tensor product of a vector $f\in\mathfrak{H}$ by $f^{\otimes i}\in \mathfrak{H}^i$, and the $i$-fold tensor product of an operator $A:\mathfrak{H}\rightarrow\mathfrak{H}$ by $A^{\otimes i}$. Denote by $\mathfrak{S}^1(\mathfrak{X})$ the space of trace-class operators on a given Hilbert space $\mathfrak{X}$ \cite{Simon1979TraceIdeals}. Define the set of states on $\mathfrak{X}$ by
	\begin{equation*}
		\mathcal{S}(\mathfrak{X}) = \left\{\Gamma \in \mathfrak{S}^1(\mathfrak{X}): \Gamma = \Gamma^* \geq 0, \Tr_\mathfrak{X}\Gamma = 1\right\}.
	\end{equation*}
	The $k$-particle reduced density matrix $\Gamma^{(k)}$ of a given state $\Gamma\in \mathcal{S}(\mathfrak{H}^N)$ is obtained by taking the partial trace over all but the first $k$ variables:
	\begin{equation*}
		\Gamma^{(k)} = \Tr_{k + 1\rightarrow N}\Gamma.
	\end{equation*}
	Given a wavefunction $\Psi\in\mfH^N$, we define its $k$-particle reduced density matrix to be that of the pure state $\vert\Psi\rangle\langle\Psi\vert$.
	
	Given a multi-index $\underline{J} = (j_1,\dots,j_M)\in\mathbb{N}_0^M$, define $\vert\underline{J}\vert = \sum_{i=1}^M j_i$. Moreover, given a wavefunction $\Psi\in\mfH^N$ and a family of mutually orthogonal projections $P_1,\dots,P_M$, define
	\begin{equation}
		\label{eq:occupancy_wavefunction_definition}
		\Psi_{\underline{J}} = P_1^{\otimes j_1}\otimes_{\textmd{s}}\dots\otimes_{\textmd{s}}P_M^{\otimes j_M}\Psi, \quad \Gamma_{\underline{J}} = \left\vert\Psi_{\underline{J}}\right\rangle\left\langle\Psi_{\underline{J}}\right\vert \quad \textmd{and} \quad \Gamma_{\underline{J},\underline{J}'} = \left\vert\Psi_{\underline{J}'}\right\rangle\left\langle\Psi_{\underline{J}}\right\vert,
	\end{equation}
	for all $\underline{J},\underline{J}'\in\mathbb{N}^M$ with $\vert\underline{J}\vert = \vert\underline{J}'\vert = N$. Furthermore, for any $i_1,i_2,i_3\in\{1,\dots,M\}$, define
	\begin{equation}
		\label{eq:multi_index_J_def_expo}
		\underline{J}^{(i_1i_2i_3)} = \left(\tilde{j}_1,\dots,\tilde{j}_M\right) \quad \textmd{with} \quad \tilde{j}_k = j_k + \delta_{i_1k} + \delta_{i_2k} + \delta_{i_3k}
	\end{equation}
	and
	\begin{equation*}
		\Gamma_{\underline{J}^{(i_1i_2i_3;i_1'i_2'i_3')}} = \Gamma_{\underline{J}^{(i_1i_2i_3)},\underline{J}^{(i_1'i_2'i_3')}}.
	\end{equation*}
	Finally, define
	\begin{equation}
		\label{eq:projections_definition}
		P_{ij} = P_i\otimes P_j, \quad P_{ijk} = P_i\otimes P_j \otimes P_k, \quad \bbP_i = \sum_{j = 1}^iP_j \quad \textmd{and} \quad \bbQ_i = 1 - \bbP_i.
	\end{equation}
	
	\subsection{Estimates on the interaction terms}
	
	We recall some estimates on the smeared Coulomb potential $w_R$, which exploit the regularising effect of the smeared charge.
	
	\begin{lemma}[Smeared Coulomb potential]
		\label{lemma:smeared_coulomb_potential}
		Let $w_R$ be defined as in \eqref{eq:smeared_coulomb_potential}. Then, there exists a universal constant $C > 0$ such that
		\begin{equation}
			\label{eq:smeared_coulomb_potential_estimates}
			\sup_{B(0,1)}\vert w_R\vert \leq C + \vert\log R\vert, \quad \sup_{\mathbb{R}^2}\vert\nabla w_R\vert \leq CR^{-1} \quad \textmd{and} \quad \sup_{B(0,1)^{\complement}}\vert \nabla w_R\vert \leq C,
		\end{equation}
		for $R$ small enough.
	\end{lemma}
	
	\begin{proof}
		See \cite[Lemma 2.1]{Lundholm2015AverageFA}.
	\end{proof}
	
	\begin{lemma}[Singular two-body term]
		\label{lemma:singular_two_body_term_bound}
		There exists a universal constant $C > 0$ such that, for $R$ small enough,
		\begin{equation}
			\label{eq:singular_two_body_term_bound}
			\vert \nabla w_R(x_1 - x_2)\vert^2 \leq C\vert\log R\vert^2\left(1 - \Delta_1\right)
		\end{equation}
		in the quadratic form sense on $L^2(\R^4)$.
	\end{lemma}
	
	\begin{proof}
		See \cite[Lemma 2.5]{Girardot2025derivationCSS}.
	\end{proof}
	
	\begin{lemma}[Three-body term]
		\label{lemma:three_body_term_bound}
		There exists a universal constant $C > 0$ such that, for $R$ small enough,
		\begin{equation}
			\label{eq:three-body_term_bound_nonneg}
			0 \leq \sum_{\substack{1\leq j,k,\ell\leq 3\\ j\neq k\neq \ell\neq j}}\nabla^\perp w_R(x_j - x_k)\cdot\nabla^\perp w_R(x_j - x_\ell) \leq C\left(1 - \Delta_1 - \Delta_2 - \Delta_3\right)
		\end{equation}
		in the quadratic form sense on $L^2(\R^6)$.
	\end{lemma}
	
	\begin{proof}
		See \cite[Lemma 2.4]{Lundholm2015AverageFA}.
	\end{proof}
	
	\subsection{Combinatorial results and plane wave estimate}
	
	In this section, we provide some combinatorial results that play a significant role in the proof of Theorem \ref{th:convergence_gse_average_field}. Also, we give an important plane wave estimate.
	
	\begin{lemma}
		\label{lemma:occupancy_wavefunction_decomposition}
		Let $\Psi\in\mfH^N$ and let $P_1,\dots,P_M$ be a family of mutually orthogonal projections on $\mathfrak{H}$ that form a resolution of the identity. Define $\Psi_{\underline{J}}$ as in \eqref{eq:occupancy_wavefunction_definition}. Then,
		\begin{equation}
			\label{eq:occupancy_wavefunction_decomposition}
			\Psi = \sum_{\underline{J}}\Psi_{\underline{J}},
		\end{equation}
		where we are summing over all $\underline{J}\in\bbN_0^M$ with $\vert\underline{J}\vert = N$. Moreover, for all $i_1,i_2,i_3\in\{1,\dots,M\}$,
		\begin{equation}
			\label{eq:occupancy_wavefunction_decomposition_with_contstraint}
			P_{i_1i_2i_3}\otimes\mathds{1}^{\otimes (N  - 3)}\Psi = \sum_{\underline{J}}P_{i_1i_2i_3}\otimes\mathds{1}^{\otimes (N  - 3)}\Psi_{\underline{J}^{(i_1i_2i_3)}},
		\end{equation}
		where we are summing over all $\underline{J}\in\bbN_0^M$ with $\vert\underline{J}\vert = N - 3$. 
	\end{lemma}
	
	\begin{proof}
		To prove \eqref{eq:occupancy_wavefunction_decomposition}, we write
		\begin{equation*}
			\Psi = (P_1 + \cdots + P_M)^{\otimes N}\Psi
		\end{equation*}
		and expand the tensor product. This leads to a sum of $M^N$ terms, which we then group according to the number of times that each projection $P_i$ occurs (counted by $j_i$). Rewriting this in terms of $\osym$ gives \eqref{eq:occupancy_wavefunction_decomposition}.
		
		Next, we use \eqref{eq:occupancy_wavefunction_decomposition} to write
		\begin{equation*}
			P_{i_1i_2i_3}\otimes\mathds{1}^{\otimes (N  - 3)}\Psi = \sum_{\underline{J}'}P_{i_1i_2i_3}\otimes\mathds{1}^{\otimes (N  - 3)}\Psi_{\underline{J}'},
		\end{equation*}
		with $\vert\underline{J}'\vert = N$. Then, we notice that the summand in the right vanishes unless $P_1^{\otimes j_1}\otimes_{\textmd{s}}\dots\otimes_{\textmd{s}}P_M^{\otimes j_M}$ contains the projections $P_{i_1}, P_{i_2}$ and $P_{i_3}$ at least once (counted with multiplicity if two indices coincide). Hence, the only multi-indices that contribute to the sum are of the form
		\begin{equation*}
			\underline{J} = \underline{J}' + (\dots,0,1,0,\dots) + (\dots,0,1,0,\dots) + (\dots,0,1,0,\dots),
		\end{equation*}
		where $\vert\underline{J}'\vert = N - 3$, and the $1$'s are in positions $i_1$, $i_2$ and $i_3$ respectively, which is precisely \eqref{eq:occupancy_wavefunction_decomposition_with_contstraint}.
	\end{proof}
	
	\begin{lemma}
		\label{lemma:state_three_projections_full_trace}
		Let $P_1,P_2,P_3$ and $Q$ be mutually orthogonal projections on $\mathfrak{H}$. Given a state
		\begin{equation*}
			\Gamma\in \mathcal{S}\big(P_1^{\otimes j_1}\otimes_\textmd{s} P_2^{\otimes j_2}\otimes_{\textmd{s}}P_3^{\otimes j_3}\otimes_\textmd{s} Q^{\otimes (N-j_1-j_2)}\mathfrak{H}^N\big),
		\end{equation*}
		for some $j_1,j_2,j_3 \geq 1,$ we have
		\begin{equation}
			\label{eq:state_one_and_two_projections_full_trace}
			\Tr\big(P_1^{\otimes 3}\Gamma^{(3)}\big) = \dfrac{j_1(j_1 - 1)(j_1 - 2)}{N(N-1)(N - 2)}, \quad \Tr\big(P_1^{\otimes 2}\otimes P_2 \Gamma^{(3)}\big) = \dfrac{j_1(j_1 - 1)j_2}{N(N-1)(N - 2)}
		\end{equation}
		and
		\begin{equation}
			\label{eq:state_three_projections_full_trace}
			\Tr\big(P_1\otimes P_2\otimes P_3 \Gamma^{(3)}\big) = \dfrac{j_1j_2j_3}{N(N-1)(N - 2)}.
		\end{equation}
	\end{lemma}
	
	\begin{proof}[Proof of Lemma~\ref{lemma:state_three_projections_full_trace}]
		The proof is a straightforward adaptation of \cite[Lemma 4]{Junge2025DerivationHT2D}.
	\end{proof}
	
	\begin{lemma}
		\label{lemma:state_one_projection_new_state}
		Let $P$ and $Q$ be mutually orthogonal projections on $\mathfrak{H}$. Given a state
		\begin{equation*}
			\Gamma \in \mathcal{S}\big(P^{\otimes j}\otimes_{\textmd{s}}Q^{\otimes (N - j)}\mathfrak{H}^N\big),
		\end{equation*}
		for some $j \geq 1$, there exists another state
		\begin{equation*}
			\Gamma_{j}\in\mathcal{S}\big(P^{\otimes j}\mathfrak{H}^j\big)
		\end{equation*}
		such that
		\begin{equation}
			\label{eq:state_one_porjection_new_state}
			P^{\otimes 3}\Gamma^{(3)}P^{\otimes 3} = {N \choose 3}^{-1}{j \choose 3}\Gamma_{j}^{(3)}.
		\end{equation}
	\end{lemma}
	
	\begin{proof}
		The proof is a straightforward adaptation of \cite[Lemma 5]{Junge2025DerivationHT2D}.
	\end{proof}
	
	\begin{proposition}[Plane wave estimate]
		\label{prop:plane_wave_estimate}
		Let $V$ satisfy \eqref{eq:trapping_potential_regularity_assumption} and define $h = -\Delta + V$. Denote by $\bbP_i$ the spectral projection $\mathds{1}(h < N^{2i\varepsilon})$, for any $i,\varepsilon > 0$. Let $\mathbf{e}_k$ be the multiplication operator on $\mathfrak{H}$ by either $\cos(k\cdot x)$ or $\sin(k\cdot x)$. Then,
		\begin{equation}
			\label{eq:plane_wave_estimate}
			\pm \bbP_i\mathbf{e}_k\bbP_i \leq \bbP_i\min\left\{1,C\dfrac{N^{2i\varepsilon}}{\vert k\vert^2}\right\}\bbP_i,
		\end{equation}
		for some universal constant $C > 0$.
	\end{proposition}
	
	The proof is the same as that of \cite[Proposition 4]{Junge2025DerivationHT2D}, and we provide it in full detail to account for the missing assumption \eqref{eq:trapping_potential_regularity_assumption}.
	
	\begin{proof}
		Take $k \in\R^2\setminus\{0\}$ and $\mathbf{e}_k = \cos(k\cdot x)$ ($\mathbf{e}_k = \sin(k\cdot x)$ being similar). The first bound in \eqref{eq:plane_wave_estimate} is just $\vert\mathbf{e}_k\vert \leq 1$. To prove the other bound, we take some smooth compactly supported function $f$ and use an integration by parts to deduce
		\begin{equation*}
			\langle \bbP_if,\cos(k\cdot x)\bbP_if\rangle = \int_{\R^2}\d{}x\vert \bbP_if\vert^2\cos(k\cdot x) = \int\d{}x(-\Delta\vert \bbP_if\vert^2)\vert k\vert^{-2}\cos(k\cdot x).
		\end{equation*}
		Here, we used the identity $\cos(k\cdot x) =-\Delta\cos(k\cdot x)/\vert k\vert^2$. Hence, an application of the Cauchy--Schwarz inequality implies
		\begin{equation*}
			\langle \bbP_if,\cos(k\cdot x)\bbP_if\rangle \leq C\vert k\vert^{-2}(\Vert \nabla \bbP_if\Vert^2 + \Vert \bbP_if\Vert\Vert\Delta \bbP_if\Vert).
		\end{equation*}
		Therefore, the claim follows from the bounds
		\begin{equation}
			\label{eq:laplacian_and_square_bdd_projections}
			\bbP_i(-\Delta)\bbP_i \leq N^{2i\varepsilon}\bbP_i \quad \textmd{and} \quad \bbP_i(-\Delta)^2\bbP_i \leq CN^{4i\varepsilon}\bbP_i
		\end{equation}
		and a density argument. The first bound is an immediate consequence of the nonnegativity of $V$. The second bound follows from the estimate
		\begin{equation*}
			(-\Delta)^2 \leq C + C(\Delta + V) + C(-\Delta + V)^2,
		\end{equation*}
		which results from \eqref{eq:trapping_potential_regularity_assumption} (see e.g. \cite[Theorem 3.23]{Lewin24}). This concludes the proof of Proposition \ref{prop:plane_wave_estimate}.
	\end{proof}
	
	\subsection{A priori energy bounds}
	
	We provide here some bounds on the ground state energy per particle $e_N^R$, and on the kinetic energy of ground states of $H_N^R$.
	
	\begin{proposition}[A priori bounds on the ground state]
		\label{prop:energy_bound_ground_state}
		Take $R$ as in Theorem~\ref{th:convergence_gse_average_field}. Then, the many-body ground state energy per particle $e_N^R$ defined in \eqref{eq:ground_state_energy_many_body} satisfies
		\begin{equation}
			\label{eq:ground_state_energy_limsup}
			\limsup_{N\rightarrow\infty}e_N^R \leq e^\af,
		\end{equation}
		with $e^\af$ given by \eqref{eq:ground_state_energy_af_singular}. Define $h$ and $\tilde{H}_3^R$ as in \eqref{eq:one_body_operator_def} and \eqref{eq:three_body_hamiltonian} respectively. Let $\Psi$ be a ground state of $H_N^R$ and denote by $\Gamma^{(k)}$ its $k$-particle reduced density matrix. Then,
		\begin{equation}
			\label{eq:kinetic_energy_bounded_a_priori}
			\Tr(h\Gamma^{(1)}) \leq C
		\end{equation}
		and
		\begin{equation}
			\label{eq:kinetic_energy_bounded_a_priori_intermediate}
			c\Tr(h\Gamma^{(1)}) + C\Tr(\tilde{H}_3^R\Gamma^{(3)}) \leq e_N.
		\end{equation}
	\end{proposition}
	
	\begin{proof}
		For $R$ of the form \eqref{eq:smearing_radius_polynomial}, the bound \eqref{eq:ground_state_energy_limsup} was already proved in \cite[Section 3.3]{Lundholm2015AverageFA} using the trial state $(u_R^\af)^{\otimes N}$, where $u_R^\af$ is a normalised minimiser of $\cE_R^\af$. Moreover, this bound easily extends to $R$ of the form \eqref{eq:smearing_radius_exponential} with $\kappa < 1/2$ using the estimate \eqref{eq:singular_two_body_term_bound}. For $\kappa \geq 1/2$, however, this estimate is no longer strong enough to control the singular two-body term. Instead, one can use a Jastrow--Dyson trial state to obtain the correct bound, as was done in \cite{Ataei2025microscopicDA}. We refer to that work for more detail.
		
		The bound on the kinetic energy \eqref{eq:kinetic_energy_bounded_a_priori} was proved in \cite[Proposition 2.6]{Lundholm2015AverageFA} for $R$ of the form \eqref{eq:smearing_radius_polynomial}. Again, this easily extends to the exponential case \eqref{eq:smearing_radius_exponential} under the condition $\kappa < 1/2$ using the estimate \eqref{eq:singular_two_body_term_bound}. However, it is unclear whether a simple proof of \eqref{eq:kinetic_energy_bounded_a_priori} exists for $\kappa \geq 1/2$, and we will need to work much harder to prove it; this is explained in the last paragraphs of Sections \ref{sec:notation_preliminaries_proof_strategy} and \ref{section:proof_main_theorem}. In particular, we will use \eqref{eq:kinetic_energy_bounded_a_priori_intermediate}.
		
		To prove \eqref{eq:kinetic_energy_bounded_a_priori_intermediate}, we start by expanding
		\begin{equation*}
			H_N^R = \sum_{j = 1}^N-\Delta_j + V(x_j) + 2\alpha\p_j\cdot\A_j^R + \alpha^2(\A_j^R)^2,
		\end{equation*}
		where $A_j^R$ is given by \eqref{eq:potential_extended}. Moreover, as a consequence of the Cauchy--Schwarz inequality, we have
		\begin{equation*}
			2\alpha\p_j\cdot\A_j^R \geq -\delta(-\Delta_j) - \delta^{-1}\alpha^2(\A_j^R)^2,
		\end{equation*}
		for all $\delta > 0$. Thus, splitting $\p_j\cdot\A_j^R$ into $(1 - \tau)\p_j\cdot\A_j^R + \tau\p_j\cdot\A_j^R,$ we deduce
		\begin{equation*}
			H_N^R \geq \sum_{j = 1}^N(1 - \tau\delta)(-\Delta_j) + V(x_j) + 2\alpha(1 - \tau)\p_j\cdot\A_j^R + (1 - \tau\delta^{-1})\alpha^2(\A_j^R)^2,
		\end{equation*}
		for all $0 < \tau < 1$. Using that $V$ is bounded from below and doing some rewriting, we get
		\begin{equation*}
			H_N^R \geq \sum_{j = 1}^N(1 - \delta)\tau h_j + (1 - \tau)(h_j + 2\alpha\p_j\cdot\A_j^R) + (1 - \delta^{-1}\tau)\alpha^2(\A_j^R)^2 - C\tau\delta N.
		\end{equation*}
		Next, we expand $(\A_j^R)^2$ to separate the three-body term from the singular two-body one. Testing on both sides with $\Psi$ and using \eqref{eq:three-body_term_bound_nonneg}, we find
		\begin{multline}
			\label{eq:hamiltonian_bdd_below_singular_radii}
			e_n \geq (1 - \delta - C(1 - \delta^{-1}))\tau\Tr\big(h\Gamma^{(1)}\big) + (1 - \tau)\Tr\big(\tilde{H}_3^R\Gamma^{(3)}\big)\\
			+ (1 - \delta^{-1}\tau)\alpha^2N^2\Tr(\vert\nabla w_R(x_1 - x_2)\vert^2\Gamma^{(2)}).
		\end{multline}
		Taking $0 < \tau < \delta < 1$, it is easy to see that
		\begin{equation*}
			1 - \delta - C(1 - \delta^{-1}) > 0 \quad \textmd{and} \quad 1 - \delta^{-1}\tau > 0.
		\end{equation*}
		Hence, the last term in \eqref{eq:hamiltonian_bdd_below_singular_radii} is nonnegative, and dropping it gives \eqref{eq:kinetic_energy_bounded_a_priori_intermediate}.
	\end{proof}
	
	\subsection{Strategy of proof}
	
	To prove Theorem~\ref{th:convergence_gse_average_field} we wish to compare the many-body ground state energy per particle $e_N$ to that of the average-field functional \eqref{eq:average_field_energy_functional}
	\begin{equation*}
		e_R^{\textmd{af}} = \inf\left\{\mathcal{E}_R^{\textmd{af}}[u]: u\in\mathfrak{H},\Vert u\Vert = 1\right\},
	\end{equation*}
	and use the convergence
	\begin{equation*}
		\lim_{R\rightarrow 0}e_R^{\textmd{af}} = e^{\textmd{af}}.
	\end{equation*}
	
	The upper bound $\limsup e_N \leq e^{\af}$ is already given in Proposition~\ref{prop:energy_bound_ground_state}. On the other hand, proving a matching lower bound requires a lot more work, and we follow the general proof strategy of \cite{Junge2025DerivationHT2D}, which itself builds on ideas from \cite{Lewin2014TheMA,Lewin2017Note2D,Nam2020ImprovedStability}.  We also use ideas from \cite{Girardot2020averageFA,Lundholm2015AverageFA}. The main idea of the proof is to use a localisation in momentum space in order to reduce the infinite-dimensional problem to multiple finite-dimensional ones. Then, for each of those finite-dimensional problems, we use the following quantitative version of the quantum de Finetti theorem \cite{Brandao2017deFinetti,Li2015QF}.
	\begin{theorem}\label{th:de_finetti}
		Given a Hilbert space $\mathfrak{X}$ of dimension $d$ and a symmetric state $\Gamma_K\in \cS(\mathfrak{X}^K)$, there exists a probability measure $\mu$ on $\cS(\mathfrak{X})$ such that
		\begin{equation*}
			\Big|\Tr A\otimes B\otimes C\Big( \Gamma_K^{(3)} - \int_{\cS(\mathfrak{X})}\gamma^{\otimes 3} d\mu(\gamma)\Big)\Big| \leq c\sqrt{\frac{\log d}{K}}\Vert A\Vert_{\op}\Vert B\Vert_{\op}\Vert C\Vert_{\op}
		\end{equation*}
		for all self-adjoint operators $A, B$ and $C$ on $\mathfrak{X}$, and for some universal constant $c > 0$.
	\end{theorem}
	
	\begin{proof}
		In \cite{Brandao2017deFinetti}, the statement was proved with $A,B,C$ replaced by quantum measurement. See \cite[Proposition 3.3]{Rougerie2020NLSagain} and \cite[Lemma 3.3]{Nam2020ImprovedStability} for an adaptation to self-adjoint operators.
	\end{proof}
	
	We wish to apply Theorem~\ref{th:de_finetti} on energy subspaces of the one-body operator $h$ defined in \eqref{eq:one_body_operator_def}. More precisely, on the subspaces defined by the spectral projections
	\begin{equation}
		\label{eq:projections_momentum_space_proof_strategy}
		P_1 = \mathds{1}(h < N^{2\varepsilon}), \quad \textmd{and} \quad P_i = \mathds{1}(N^{2(i-1)\varepsilon} \leq h < N^{2i\varepsilon}), \quad 2 \leq  i \leq M,
	\end{equation}
	for some nonnegative parameters $\varepsilon$ and $M$ taken such that
	\begin{equation}
		\label{eq:proof_strategy_relation_M_epsilon}
		N^{M\varepsilon} \gg R^{-1}.
	\end{equation}
	The use of multiple projections is an important difference compared to \cite{Girardot2020averageFA} and \cite{Lundholm2015AverageFA}, where only projection was used, namely $\mathds{1}(h < R^{-2})$. In both works, a lot of effort goes into estimating the terms containing the projection $\mathds{1}(h \geq R^{-2})$, and bounding the errors coming from the application of the de Finetti theorem on $\mathds{1}(h < R^{-2})\mathfrak{H}$. In contrast, the decomposition we use here is much finer and allows us to use the kinetic energy\footnote{We often refer to the expectation value of $h$ as the kinetic energy.} much more precisely in order to control the error terms. The drawback is that working with this decomposition requires a lot more effort on the combinatorial side.
	
	Let $\Psi$ be a ground state of $H_N^R$ and denote by $\Gamma^{(k)}$ its $k$-particle reduced density matrix. Before applying Theorem \ref{th:de_finetti}, we do a lot of rewriting on the energy of $\Psi$. First, we will bound the energy of $\Psi$ from below by
	\begin{equation*}
		\langle\Psi,H_N^R\Psi\rangle \gtrsim N\Tr\big(\tilde{H}_3^R\Gamma^{(3)}\big),
	\end{equation*}
	where $\tilde{H}_3^R$ is the three-body Hamiltonian defined in \eqref{eq:three_body_hamiltonian}. Using that the spectral projections defined in \eqref{eq:projections_momentum_space_proof_strategy} form a resolution of the identity we will then roughly get
	\begin{equation*}
		\Tr\big(\tilde{H}_3^R\Gamma^{(3)}\big) \gtrsim \sum_{\substack{1\leq i_1,i_2,i_3\leq M\\ 1\leq i_1',i_2',i_3'\leq M}}\Tr\big(P_{i_1i_2i_3}\tilde{H}_3^RP_{i_1'i_2'i_3'}\Gamma^{(3)}\big).
	\end{equation*}
	Recall that $P_{i_1i_2i_3}$ was defined in \eqref{eq:projections_definition}. To discard the terms containing the projection $\mathds{1}(N^{2M\varepsilon} \leq h)$, we use that, thanks to \eqref{eq:proof_strategy_relation_M_epsilon} and the estimates \eqref{eq:smeared_coulomb_potential_estimates} and \eqref{eq:three-body_term_bound_nonneg}, the contribution of $h$ largely controls that of the interaction terms on the subspace $\mathds{1}(N^{2M\varepsilon} \leq h)\mathfrak{H}$.
	
	Next, we will decompose the many-body wavefunction according to the occupancy of its energy levels. Namely, we decompose $\Psi$ as in \eqref{eq:occupancy_wavefunction_decomposition} to deduce
	\begin{equation}
		\label{eq:hamiltonian_lower_bound_strategy_proof_first}
		\Tr\big(\tilde{H}_3^R\Gamma^{(3)}\big) \gtrsim \sum_{\underline{J},\underline{J}'}\sum_{\substack{1\leq i_1,i_2,i_3\leq M\\ 1\leq i_1',i_2',i_3'\leq M}}\Tr\big(P_{i_1i_2i_3}\tilde{H}_3^RP_{i_1'i_2'i_3'}\Gamma_{\underline{J},\underline{J}'}^{(3)}\big),
	\end{equation}
	where the sum is taken over $\underline{J},\underline{J}'\in\bbN_0^{M+1}$ with $\vert\underline{J}\vert = \vert\underline{J}'\vert = N$.	After that, we define
	\begin{equation}
		\label{eq:imax_def_poly}
		\imax(\underline{J}) = \max\big\{i\in\{1,\dots,M\}:j_i \geq N^{1 - \delta\varepsilon}\big\}
	\end{equation}
	for some $\delta > 0$, and distinguish between the terms that satisfy $\imax(\underline{J}) = \imax(\underline{J}')$ and for which $i_1,\dots, i_3'$ are all less than $\imax(\underline{J})$, and the ones that do not. In the latter case, we will show that their energy is much smaller than the overall kinetic energy of the system and that they can therefore be neglected. Furthermore, it is crucial to discard these terms \textit{before} applying Theorem \ref{th:de_finetti}, because they produce large errors. Once these terms are neglected, we are left with
	\begin{equation}
		\label{eq:hamiltonian_lower_bound_strategy_proof_second}
		\Tr\big(\tilde{H}_3^R\Gamma^{(3)}\big) \gtrsim \sum_{\substack{\underline{J},\underline{J}'\\ \imax(\underline{J}) = \imax(\underline{J}')}}\sum_{\substack{i_1,i_2,i_3 = 1\\ i_1',i_2',i_3' = 1}}^{\imax(\underline{J})}\Tr\big(P_{i_1i_2i_3}\tilde{H}_3^RP_{i_1'i_2'i_3'}\Gamma_{\underline{J},\underline{J}'}^{(3)}\big).
	\end{equation}
	Going from \eqref{eq:hamiltonian_lower_bound_strategy_proof_first} to \eqref{eq:hamiltonian_lower_bound_strategy_proof_second} is one of the main difficulties in adapting the proof strategy of \cite{Junge2025DerivationHT2D} and, in particular, new ideas are required to deal with the momentum operators contained in the potential $W_2$. Since the proof of \eqref{eq:hamiltonian_lower_bound_strategy_proof_second} is the key technical part of the paper, we discuss it briefly. A naive application of the Cauchy--Schwarz inequality shows that the terms we wish to control can be bounded by
	\begin{equation*}
		\Tr\big(P_{i_1i_2i_3}W_3P_{i_1i_2i_3}\Gamma_{\underline{J}}^{(3)}\big),
	\end{equation*}
	where one the indices, say $i_3$, is such that $j_{i_3} \leq N^{1 - \delta\varepsilon}$ (and likewise for $W_2$). Then, an application of Lemmas \ref{lemma:three_body_term_bound} and \ref{lemma:state_three_projections_full_trace} roughly implies
	\begin{equation*}
		\Tr\big(P_{i_1i_2i_3}W_3P_{i_1i_2i_3}\Gamma_{\underline{J}}^{(3)}\big) \lesssim N^{-\delta\varepsilon}\Tr\big(P_{i_1}h_1P_{i_1}\Gamma_{\underline{J}}^{(1)}\big) \ll \Tr(h\Gamma^{(1)}),
	\end{equation*}
	meaning that we can hope to show that these terms are controlled by the kinetic energy. Though this turns out to be the case---and is the statement of Lemma \ref{lemma:low_occupancy_estimates} below---we encounter several important technical difficulties when trying to prove it, and therefore postpone the proof to Section \ref{section:proof_main_estimates}.
	
	Fix $i = \imax(\underline{J})$ in \eqref{eq:hamiltonian_lower_bound_strategy_proof_second} and write
	\begin{equation*}
		\sum_{\substack{i_1,i_2,i_3 = 1\\ i_1',i_2',i_3' = 1}}^i\Tr\big(P_{i_1}\otimes P_{i_2}\otimes P_{i_3}\tilde{H}_3^RP_{i_1'}\otimes P_{i_2'}\otimes P_{i_3'}\Gamma_{\underline{J},\underline{J}'}^{(3)}\big) = \Tr\big(\bbP_i^{\otimes 3}\tilde{H}_3^R\bbP_i^{\otimes 3}\Gamma_{\underline{J},\underline{J}'}^{(3)}\big).
	\end{equation*}
	Recall that $\bbP_j$ was defined in \eqref{eq:projections_definition}. This is almost of the correct form to apply Theorem~\ref{th:de_finetti} on the space $\mathfrak{X} = \bbP_i\mathfrak{H}$. The only remaining issue is that we can only apply Theorem~\ref{th:de_finetti} to a state belonging to $\mathcal{S}(\mathfrak{X}^K)$, for some $K$, and we currently have $\Gamma_{\underline{J},\underline{J}'}$, which is in general not even a state. Said differently, we would like to know exactly how many particle have momenta in $\bbP_i$, and discard the information about the rest. This can be done by fixing, in addition to $i = \imax(\underline{J})$, the number $K$ of particles that have momenta in $\bbP_i$, and using Lemma~\ref{lemma:state_one_projection_new_state} to construct a state $\gamma_{i,K}\in\mathcal{S}(\bbP_i^{\otimes K}\mathfrak{H}^K)$. Having done so, we will obtain
	\begin{equation}
		\label{eq:hamiltonian_lower_bound_strategy_proof}
		\Tr\big(\tilde{H}_3^R\Gamma^{(3)}\big) \gtrsim \sum_{i = 1}^M\sum_K{N \choose 3}^{-1}{K \choose 3}\Vert\Psi_{i,K}\Vert^2\Tr\big(\tilde{H}_3^R\gamma_{i,K}^{(3)}\big),
	\end{equation}
	where $\Psi_{i,K}$ is given by
	\begin{equation*}
		\Psi_{i,K} = \sum_{\substack{\underline{J}\\ \imax(\underline{J}) = i\\ \jmax = K}}\Psi_{\underline{J}}, \quad \textmd{with} \quad \jmax = \sum_{k = 1}^{\imax(\underline{J})}j_k.
	\end{equation*}
	For each state $\gamma_{i,K}$, we will apply Theorem~\ref{th:de_finetti} using the CLR-type estimate \eqref{eq:clr_estimate} and Proposition~\ref{prop:plane_wave_estimate} to bound the error terms, to obtain
	\begin{equation}
		\label{eq:application_de_finetti_strategy_proof}
		\Tr\big(\tilde{H}_3^R\gamma_{i,K}^{(3)}\big) \gtrsim \int_{\mathcal{S}(\bbP_i\mathfrak{H})}\d{}\mu(\gamma)\Tr\big(\tilde{H}_3^R\gamma^{\otimes 3}\big) \gtrsim e_R^{\textmd{af}}.
	\end{equation}
	Inserting \eqref{eq:application_de_finetti_strategy_proof} into \eqref{eq:hamiltonian_lower_bound_strategy_proof} and using that $e_R^\af$ converges to $e^\af$ will yield the desired result.
	
	Let us mention that, although we did not discuss it here, the bound \eqref{eq:kinetic_energy_bounded_a_priori} plays an important role in \eqref{eq:application_de_finetti_strategy_proof} because it allows us to prove that the measure $\mu$ is supported over pure states. However, up to this point we have not proved this bound for $R$ of the form \eqref{eq:smearing_radius_exponential} with $1/2 \leq \kappa < 1$. To solve this issue, we first use the much cruder bound $\Tr(\tilde{H}_3^R\gamma^{\otimes 3}) \geq -C$ and the estimate \eqref{eq:kinetic_energy_bounded_a_priori_intermediate} to deduce that \eqref{eq:kinetic_energy_bounded_a_priori} is true for $1/2 \leq \kappa < 1$, and after that we conclude the proof as in the other cases.
	
	\section{Mean-field limit: proof of Theorem~\ref{th:convergence_gse_average_field}}
	
	\label{section:proof_main_theorem}
	
	We establish the lower bound
	\begin{equation}
		\label{eq:energy_main_bound}
		e_N^R \geq e_R^{\textmd{af}} - CN^{-\nu}
	\end{equation}
	for some $\nu > 0$, and where $e_R^{\textmd{af}}$ denotes the ground state energy of the functional $\cE_R^\af$ defined in \eqref{eq:average_field_energy_functional}. Then, the energy convergence \eqref{eq:energy_converence_main_result} follows from
	\begin{equation*}
		e_R^{\textmd{af}} \longrightarrow e^{\textmd{af}}
	\end{equation*}
	as $N \rightarrow \infty$ (see \cite[Appendix A]{Girardot2020averageFA}). The state convergence follows directly using ideas from \eqref{eq:state_convergence_main_result} \cite{Lewin2014TheMA} (see also \cite{Girardot2020averageFA,Lundholm2015AverageFA}). The upper bound matching \eqref{eq:energy_main_bound} is given in Proposition~\ref{prop:energy_bound_ground_state}.
	
	Let $\Psi$ be a ground state of $H_N^R$ and denote by $\Gamma$ the projection $\vert\Psi\rangle\langle\Psi\vert$. Define the operators $T, W_2$ and $W_3$ as is \eqref{eq:effective_hamiltonian_three_terms_def}. By symmetry of $\Psi$, the nonnegativity of the fourth term in \eqref{eq:hamiltonian_regularised_rewritten}, and the assumption \eqref{eq:statistics_parameter_main_theorem} on $\alpha$, we have
	\begin{align*}
		N^{-1}\left\langle\Psi,H_N^R\Psi\right\rangle &\geq \Tr\big(T\Gamma^{(3)}\big) + \beta\Tr\big(W_2\Gamma^{(3)}\big) + \beta^2\dfrac{N - 2}{N - 1}\Tr\big(W_3\Gamma^{(3)}\big).
	\end{align*}
	The factor $(N - 2)/(N - 1)$ in the right can be replaced by $1$ at the cost of a small amount of the kinetic energy. Specifically, Lemma~\ref{lemma:three_body_term_bound} implies
	\begin{equation*}
		\beta^2\left\vert1 - \dfrac{N - 2}{N - 1}\right\vert\big\vert\Tr\big(W_3\Gamma^{(3)}\big)\big\vert \leq CN^{-1}\Tr\big((C + h)\Gamma^{(1)}\big).
	\end{equation*}
	Hence, by defining the effective $\tilde{H}_3^R$ as in \eqref{eq:three_body_hamiltonian}, we obtain
	\begin{equation}
		N^{-1}\left\langle\Psi,H_N^R\Psi\right\rangle \geq \Tr\big(\tilde{H}_3^{R}\Gamma^{(3)}\big) - CN^{-1}\Tr\big((C + h)\Gamma^{(1)}\big). \label{eq:hamiltonian_no_high_momenta_bound}
	\end{equation}
	
	Next, we split $\mathfrak{H}$ into $M + 1$ annuli in momentum space according to the spectral projections
	\begin{equation}
		\label{eq:projections_momentum_space}
		\begin{array}{c}
			P_1 = \mathds{1}(h < N^{2\varepsilon}), \quad P_i = \mathds{1}(N^{2(i-1)\varepsilon} \leq h < N^{2i\varepsilon}), \quad 2 \leq  i \leq M,\\
			P_{M + 1} = \mathds{1}(N^{2M\varepsilon}\leq h),
		\end{array}
	\end{equation}
	for some nonnegative parameters $\varepsilon$ and  $M$ that will be chosen later. In the polynomial case $R = N^{-\eta}$, the parameter $\varepsilon$ will be universal, and in the exponential case $R = e^{-N^\kappa}$ it will depend only on $\kappa$. The parameter $M$ shall be taken such that
	\begin{equation}
		\label{eq:relation_M_epsilon}
		M = \mathcal{O}\left(\log R^{-1}/\log N\right).
	\end{equation}
	In particular, $M$ will be constant in the polynomial case. Thanks to assumption \eqref{eq:trapping_potential_assumption} and \eqref{eq:relation_M_epsilon}, we have the Cwikel--Lieb--Rosenblum (CLR) type estimate
	\begin{equation}
		\label{eq:clr_estimate}
		\dim(\mathbb{P}_M\mathfrak{H}) \leq CR^{-C}.
	\end{equation}
	(see \cite[Lemma 3.3]{Lewin2014TheMA} and references therein).
	
	We claim that
	\begin{equation*}
		N^{-1}\left\langle\Psi,H_N^R\Psi\right\rangle \geq \sum_{\substack{i_1,i_2,i_3 = 1\\ i_1',i_2',i_3' = 1}}^M\Tr\big(P_{i_1i_2i_3}\tilde{H}_3^RP_{i_1'i_2'i_3'}\Gamma^{(3)}\big) - CN^{-2\varepsilon}\Tr\big((C + h)\Gamma^{(1)}\big) \numberthis \label{eq:hamiltonian_three_body_effective_hamiltonian_bound}
	\end{equation*}
	(taking $\varepsilon$ small enough in \eqref{eq:projections_momentum_space}). To prove this, we use the resolution of the identity
	\begin{equation}
		\label{eq:resolution_identity}
		\sum_{i=1}^{M + 1} P_i = \mathds{1}
	\end{equation}
	to expand
	\begin{equation*}
		\Tr\big(\tilde{H}_3^R\Gamma^{(3)}\big) = \sum_{\substack{i_1,i_2,i_3 = 1\\ i_1',i_2',i_3' = 1}}^{M + 1}\Tr\big(P_{i_1i_2i_3}\tilde{H}_3^RP_{i_1'i_2'i_3'}\Gamma^{(3)}\big).
	\end{equation*}
	After that, we bound the interaction terms containing at least one projection $P_{M + 1}$ using Lemma~\ref{lemma:high_momenta_estimates}, and insert the result into \eqref{eq:hamiltonian_no_high_momenta_bound} to deduce \eqref{eq:hamiltonian_three_body_effective_hamiltonian_bound}. An example of such a term is
	\begin{equation*}
		\sum_{\substack{i_2,i_3\\ i_1',i_2',i_3'}}^M\Tr\big(P_{(M+1)i_2i_3}\tilde{H}_3^RP_{i_1'i_2'i_3'}\Gamma^{(3)}\big) = \Tr\big(P_{M+1}\otimes\bbP_M\otimes\bbP_M\tilde{H}_3^R\bbP_M\otimes\bbP_M\otimes\bbP_M\Gamma^{(3)}\big),
	\end{equation*}
	which we rewrote using the projection $\bbP_M$ defined in \eqref{eq:projections_definition}; all the others can be rewritten in a similar way. Since Lemma~\ref{lemma:high_momenta_estimates} will not be used again in the proof of Theorem \ref{th:convergence_gse_average_field}, we postpone its own proof to Section \ref{section:proof_main_estimates}.
	
	\begin{lemma}[High-momenta estimates]
		\label{lemma:high_momenta_estimates}
		Take $R$ as in Theorem~\ref{th:convergence_gse_average_field}, and $M$ satisfying \eqref{eq:relation_M_epsilon}. Then, there exists a universal constant $C > 0$ such that, for any state $\Gamma\in\cS(\fH^3)$, and any projections $Q_1,\dots,Q_5\in \{\bbP_M,P_{M + 1}\}$,
		\begin{equation}
			\label{eq:high_momenta_estimate_W2}
			\left\vert\Tr\left(P_{M + 1}\otimes Q_1 \otimes Q_2 W_2 Q_3\otimes Q_4 \otimes Q_5\Gamma\right)\right\vert \leq CN^{-2\varepsilon}\Tr\big((C + h)\Gamma^{(1)}\big)
		\end{equation}
		and
		\begin{equation}
			\label{eq:high_momenta_estimate_W3}
			\left\vert\Tr\left(P_{M + 1}\otimes Q_1 \otimes Q_2 W_3 Q_3\otimes Q_4 \otimes Q_5\Gamma\right)\right\vert \leq CN^{-2\varepsilon}\Tr\big((C + h)\Gamma^{(1)}\big).
		\end{equation}
	\end{lemma}
	
	Next, we decompose $\Psi$ according to the occupation of each energy level $P_i$. Namely, we decompose $\Psi$ as in \eqref{eq:occupancy_wavefunction_decomposition_with_contstraint}, which yields the identity
	\begin{equation*}
		P_{i_1'i_2'i_3'}\Gamma^{(3)}P_{i_1i_2i_3} = \sum_{\underline{J}}P_{i_1'i_2'i_3'}\Gamma_{\underline{J}^{(i_1i_2i_3;i_1'i_2'i_3')}}^{(3)}P_{i_1i_2i_3}.
	\end{equation*}
	Injecting this into \eqref{eq:hamiltonian_three_body_effective_hamiltonian_bound}, we find
	\begin{equation}
		\label{eq:three_body_effective_hamiltonian_proj_decomposition}
		N^{-1}\langle\Psi,H_N^R\Psi\rangle \geq \sum_{\underline{J}}\sum_{\substack{i_1,i_2,i_3 = 1\\ i_1',i_2',i_3' = 1}}^M\Tr\big(P_{i_1i_2i_3}\tilde{H}_3^RP_{i_1'i_2'i_3'}\Gamma_{\underline{J}^{(i_1i_2i_3;i_1'i_2'i_3')}}^{(3)}\big) - CN^{-2\varepsilon}\Tr\big((C + h)\Gamma^{(1)}\big),
	\end{equation}
	where we are summing over all multi-indices $\underline{J}\in\bbN_0^{M+1}$ such that $\vert\underline{J}\vert = N - 3$.
	
	To bound the first term in the right-hand side of \eqref{eq:three_body_effective_hamiltonian_proj_decomposition}, we define $\imax(\underline{J})$ as in \eqref{eq:imax_def_poly} for some $\delta > 0$ to be chosen later. For readability's sake, we sometimes omit $\underline{J}$ and simply write $\imax$. Also, by convention, we take $\imax(\underline{J}) = 0$ if the set on the right is empty. We make the following important distinction regarding the sum over $i_1,\dots,i_3'$ in \eqref{eq:three_body_effective_hamiltonian_proj_decomposition}: either all indices are less than $\imax$---in which case we use Theorem~\ref{th:de_finetti}---or not. In the latter case, we need to bound
	\begin{equation*}
		\Tr\big(P_{i_1i_2i_3}W_2P_{i_1'i_2'i_3'}\Gamma_{\underline{J}^{(i_1i_2i_3;i_1'i_2'i_3')}}^{(3)}\big) \quad \textmd{and} \quad \Tr\big(P_{i_1i_2i_3}W_3P_{i_1'i_2'i_3'}\Gamma_{\underline{J}^{(i_1i_2i_3;i_1'i_2'i_3')}}^{(3)}\big)
	\end{equation*}
	when one the indices, say $i_1$, is such that $j_{i_1} < N^{1 - \delta\varepsilon}$. This is done in Lemma~\ref{lemma:low_occupancy_estimates}, which, from the technical point of view, is the main novelty of the present work. Since the proof is rather complicated and intricate, we postpone it to Section \ref{section:proof_main_estimates}.
	
	\begin{lemma}[Low-occupancy estimates]
		\label{lemma:low_occupancy_estimates}
		Take $R$ as in Theorem~\ref{th:convergence_gse_average_field}, and $M$ satisfying \eqref{eq:relation_M_epsilon}. Define $\imax$ as in \eqref{eq:imax_def_poly}. Then, for all $\Lambda > 0$,
		\begin{multline}
			\label{eq:low_momenta_estimate_W2}
			\Big\vert\sum_{\underline{J}}\sum_{i_1 = \imax + 1}^M\sum_{i_2,i_3}\sum_{i_1',i_2',i_3'}\Tr\big(P_{i_1i_2i_3}W_2P_{i_1'i_2'i_3'}\Gamma_{\underline{J}^{(i_1i_2i_3;i_1'i_2'i_3')}}^{(3)}\big)\Big\vert\\
			\leq C_\Lambda \big(N^{-(\Lambda - 3)\varepsilon/2} + N^{-(\delta/2 - 2)\varepsilon} + N^{-(1 - \kappa)/2 + 2\varepsilon}\big)\Tr\big(\left(C + h\right)\Gamma^{(1)}\big)
		\end{multline}
		and
		\begin{multline}
			\label{eq:low_momenta_estimate_W3}
			\Big\vert\sum_{\underline{J}}\sum_{i_1 = \imax + 1}^M\sum_{i_2,i_3}\sum_{i_1',i_2',i_3'}\Tr\big(P_{i_1i_2i_3}W_3P_{i_1'i_2'i_3'}\Gamma_{\underline{J}^{(i_1i_2i_3;i_1'i_2'i_3')}}^{(3)}\big)\Big\vert\\
			\leq C_\Lambda \big(N^{-(\Lambda - 3)\varepsilon/2} + N^{-(\delta/2 - 2)\varepsilon} + N^{-(1 - \kappa)/2 + 2\varepsilon}\big)\Tr\big(\left(C + h\right)\Gamma^{(1)}\big),
		\end{multline}
		for some constant $C_\Lambda > 0$ depending only on $\Lambda$, and with $\kappa = 0$ in the polynomial case \eqref{eq:smearing_radius_polynomial}. Here, we are summing over all multi-indices $\underline{J}\in\bbN_0^{M+1}$ satisfying $\vert \underline{J}\vert = N - 3$.
	\end{lemma}
	
	Taking $\delta,\Lambda\geq 5$ and inserting \eqref{eq:low_momenta_estimate_W2} and \eqref{eq:low_momenta_estimate_W3} into \eqref{eq:three_body_effective_hamiltonian_proj_decomposition}, we obtain
	\begin{multline}
		N^{-1}\langle\Psi,H_N^R\Psi\rangle \geq \sum_{\underline{J}}\sum_{\substack{i_1,i_2,i_3 = 1\\ i_1',i_2',i_3' = 1}}^{\imax}\Tr\big(P_{i_1i_2i_3}\tilde{H}_3^RP_{i_1'i_2'i_3'}\Gamma_{\underline{J}^{(i_1i_2i_3;i_1'i_2'i_3')}}^{(3)}\big)\\
		- C(N^{-\varepsilon} - CN^{-(1 - \kappa)/2 + 2\varepsilon})\Tr\big((C + h)\Gamma^{(1)}\big), \label{eq:low_occupancy_estimate_polynomial}
	\end{multline}
	with $\kappa = 0$ in the polynomial case \eqref{eq:smearing_radius_polynomial}. Hence, we will take
	\begin{equation}
		\label{eq:condition_epsilon_less_kappa}
		\varepsilon < \dfrac{1 - \kappa}{4}
	\end{equation}
	to make sure that the last term in \eqref{eq:low_occupancy_estimate_polynomial} is small.
	
	Before we apply Theorem~\ref{th:de_finetti}, we do some rewriting. We decompose first over the value of $\imax$:
	\begin{equation*}
		\sum_{\underline{J}}\sum_{\substack{i_1,i_2,i_3 = 1\\ i_1',i_2',i_3' = 1}}^{\imax}P_{i_1'i_2'i_3'}\Gamma_{\underline{J}^{(i_1i_2i_3;i_1'i_2'i_3')}}^{(3)}P_{i_1i_2i_3} = \sum_{i = 1}^M\sum_{\substack{i_1,i_2,i_3 = 1\\ i_1',i_2',i_3' = 1}}^i\sum_{\substack{\underline{J}\\ \imax(\underline{J}) = i}}P_{i_1'i_2'i_3'}\Gamma_{\underline{J}^{(i_1i_2i_3;i_1'i_2'i_3')}}^{(3)}P_{i_1i_2i_3}.
	\end{equation*}
	At fixed $i_1,\dots,i_3'$, we use Lemma \ref{lemma:occupancy_wavefunction_decomposition} to rewrite the last sum in the right as
	\begin{equation}
		\label{eq:sum_J_rewritten}
		\sum_{\substack{\underline{J}\\ \imax(\underline{J}) = i}}P_{i_1'i_2'i_3'}\Gamma_{\underline{J}^{(i_1i_2i_3;i_1'i_2'i_3')}}^{(3)}P_{i_1i_2i_3} = \sum_{\substack{\underline{J}',\underline{J}''\\ \imax(\underline{J}') = i\\ \imax(\underline{J}'') = i}}P_{i_1'i_2'i_3'}\Gamma_{\underline{J}',\underline{J}''}^{(3)}P_{i_1i_2i_3},
	\end{equation}
	where we are summing over $\underline{J}$'s satisfying $\vert\underline{J}\vert = N - 3$ in the left, and over $\underline{J}',\underline{J}''$ satisfying $\vert\underline{J}'\vert = \vert\underline{J}''\vert = N$ in the right.
	
	Define now
	\begin{equation}
		\label{eq:wavefunction_decomposed_I_K}
		\Psi_{i,K} = \sum_{\substack{\underline{J}\\ \imax(\underline{J}) = i\\ \jmax(\underline{J}) = K}}\Psi_{\underline{J}} \quad \textmd{and} \quad \Gamma_{i,K} = \left\vert\Psi_{i,K}\right\rangle\left\langle\Psi_{i,K}\right\vert, \quad \textmd{with} \quad \jmax(\underline{J}) = \sum_{i = 1}^{\imax(\underline{J})}j_i.
	\end{equation}
	Thus, we can rewrite \eqref{eq:sum_J_rewritten} further as
	\begin{equation*}
		\sum_{\substack{\underline{J}\\ \imax(\underline{J}) = i}}P_{i_1'i_2'i_3'}\Gamma_{\underline{J}^{(i_1i_2i_3)},\underline{J}^{(i_1'i_2'i_3')}}^{(3)}P_{i_1i_2i_3} = \sum_KP_{i_1'i_2'i_3'}\Gamma_{i,K}^{(3)}P_{i_1i_2i_3}.
	\end{equation*}
	All in all, we have shown that
	\begin{equation*}
		N^{-1}\langle\Psi,H_N^R\Psi\rangle \geq \sum_{i = 1}^M\sum_K\Tr\big(\bbP_i^{\otimes 3}\tilde{H}_3^R\bbP_i^{\otimes 3}\Gamma_{i,K}^{(3)}\big) - C(N^{-\varepsilon} + N^{-(1 - \kappa)/2 + 2\varepsilon})\Tr\big((C + h)\Gamma^{(1)}\big), \numberthis \label{eq:low_occupancy_estimate_rewritten}
	\end{equation*}
	where we are summing over $N - MN^{1 - \delta\varepsilon} \leq K \leq N$. The parameter $\kappa$ in \eqref{eq:low_occupancy_estimate_rewritten} is equal to $0$ in the polynomial case \eqref{eq:smearing_radius_polynomial}.
	
	Since $\Psi_{i,K}$ is symmetric and belongs to $\bbP_i^{\otimes K}\otimes_{\textmd{s}}\bbQ_i^{\otimes(N - K)}\mathfrak{H}^N$, we can use Lemma~\ref{lemma:state_one_projection_new_state} to find a symmetric state $\gamma_{i,K}\in\mathcal{S}(\bbP_i^{\otimes K}\mathfrak{H}^K)$ such that
	\begin{equation*}
		\bbP_i^{\otimes 3}\Gamma_{i,K}^{(3)}\bbP_i^{\otimes 3} = {N \choose 3}^{-1}{K \choose 3}\Tr(\Gamma_{i,K})\gamma_{i,K}^{(3)}.
	\end{equation*}
	Hence, we may rewrite \eqref{eq:low_occupancy_estimate_rewritten} as
	\begin{multline}
		N^{-1}\left\langle\Psi,H_N^R\Psi\right\rangle \geq \sum_{i = 1}^M\sum_K{N \choose 3}{K \choose 3}^{-1}\Tr(\Gamma_{i,K})\Tr\big(\tilde{H}_3^R\gamma_{i,K}^{(3)}\big)\\
		- C(N^{-2\varepsilon} + N^{-(1 - \kappa)/2 + 2\varepsilon})\Tr\big((C + h)\Gamma^{(1)}\big). \label{eq:hamiltonian_lower_bound_proj_decomposition_high_occupancy}
	\end{multline}
	
	We now use Theorem~\ref{th:de_finetti} to bound $\Tr(\tilde{H}_3^R\gamma_{i,K}^{(3)})$ from below. To that end, we define
	\begin{equation*}
		\Delta\gamma_{i,K}^{(3)} = \gamma_{i,K}^{(3)} - \int_{\mathcal{S}(\bbP_i\mathfrak{H})}\gamma^{\otimes 3}\d{}\mu_{i,K}(\gamma),
	\end{equation*}
	where $\mu_{i,K}$ is the probability measure from Theorem~\ref{th:de_finetti} applied to the state $\gamma_{i,K}\in\mathcal{S}(\bbP_i^K\mathfrak{H}^K)$. Thus,
	\begin{equation}
		\label{eq:hamiltonian_lower_bound_before_de_finetti}
		\Tr\big(\tilde{H}_3^R\gamma_{i,K}^{(3)}\big) = \int_{\mathcal{S}(\bbP_i\mathfrak{H})}\Tr\big(\tilde{H}_3^R\gamma^{\otimes 3}\big)\d{}\mu_{i,K}(\gamma) + \Tr\big(\tilde{H}_3^R\Delta\gamma_{i,K}^{(3)}\big).
	\end{equation}
	Only the first term contributes to the ground state energy; the second is an error, as we now show.
	\\
	
	\noindent
	\textbf{Analysis of $T$.} A direct application of Theorem~\ref{th:de_finetti} yields
	\begin{equation}
		\label{eq:remainder_de_finetti_kinetic}
		\Tr\big(T\Delta\gamma_{i,K}^{(3)}\big) \geq -C\sqrt{\dfrac{\log R^{-1}}{K}}N^{2i\varepsilon},
	\end{equation}
	where we used the CLR-type estimate \eqref{eq:clr_estimate} to bound the dimension of $\bbP_i\mathfrak{H}$.
	\\
	
	\noindent
	\textbf{Analysis of $W_2$.} In order to apply Theorem~\ref{th:de_finetti}, we need to write $W_2$ in a tensorised form. By symmetry, it suffices to consider
	\begin{equation*}
		W_2(1,2) = \p_1\cdot\nabla^\perp w_R(x_1 - x_2).
	\end{equation*}
	For readability's sake, we define
	\begin{equation*}
		\chi_R = \dfrac{\mathds{1}_{B(0,R)}}{\pi R^2}.
	\end{equation*}
	Then, using the Fourier transform, we write
	\begin{equation*}
		W_2(1,2) = -2\pi\mathbf{i}\int\d{}k\hat{\chi}_R(k)\dfrac{e_{k^\perp}}{\vert k\vert}e^{-\mathbf{i}k\cdot x_2}\p_1 e^{\mathbf{i}k\cdot x_1},
	\end{equation*}
	where $\e_{k^\perp}$ denotes the normalised direction of the vector $k^\perp$. Moreover, using that
	\begin{equation*}
		\mathbf{p}_1 e^{\mathbf{i}k\cdot x_1} = e^{\mathbf{i}k\cdot x_1}(\mathbf{i}k + \mathbf{p}_1)
	\end{equation*}
	and that $e_{k^\perp}\cdot k = 0$, we get
	\begin{equation*}
		W_2(1,2) = -2\pi\mathbf{i}\int\d{}k\dfrac{\hat{\chi}_R(k)}{\vert k\vert}e^{\mathbf{i}k\cdot (x_1 - x_2)}e_{k^\perp}\cdot\mathbf{p}_1.
	\end{equation*}
	Developing $e^{\mathbf{i}k\cdot(x_1 - x_2)}$ and using that $k \mapsto \frac{\hat{\chi}_R(k)}{\vert k\vert}e_{k^\perp}$ is odd, we further rewrite this as
	\begin{equation}
		\label{eq:two_body_mixed_potential_fourier}
		W_2(1,2) = 2\pi\int\d{}k\dfrac{\hat{\chi}_R(k)}{\vert k\vert}\left(\sin(k\cdot x_1)\cos(k\cdot x_2) - \cos(k\cdot x_1)\sin(k\cdot x_2)\right)e_{k^\perp}\cdot\mathbf{p}_1.
	\end{equation}
	Finally, splitting the integral over $k$ between $\vert k\vert \leq N^{i\varepsilon}$ and $\vert k\vert > N^{i\varepsilon}$, and using Theorem~\ref{th:de_finetti} and Proposition~\ref{prop:plane_wave_estimate}, we obtain
	\begin{align*}
		\Tr\big(W_2\Delta\gamma_{i,K}^{(3)}\big) &\geq -C\sqrt{\dfrac{\log R^{-1}}{K}}\int_{\vert k\vert \leq N^{i\varepsilon}}\d{}k\Vert \bbP_ie_{k^\perp}\cdot\p \bbP_i\Vert_{\textmd{op}}\dfrac{\vert\hat{\chi}_R(k)\vert}{\vert k\vert}\\
		&\phantom{\geq} -C\sqrt{\dfrac{\log R^{-1}}{K}}N^{2i\varepsilon}\int_{\vert k\vert > N^{i\varepsilon}}\d{}k\Vert \bbP_ie_{k^\perp}\cdot\p \bbP_i\Vert_{\textmd{op}}\dfrac{\vert\hat{\chi}_R(k)\vert}{\vert k\vert^3}\\
		&\geq -C\sqrt{\dfrac{\log R^{-1}}{K}}N^{2i\varepsilon}. \numberthis \label{eq:remainder_de_finetti_W2}
	\end{align*}
	Here, we used that $\Vert\hat{\chi}_R\Vert_{L^\infty} \leq \Vert\chi_R\Vert_{L^1}\leq C$.
	\\
	
	\noindent
	\textbf{Analysis of $W_3$.} By symmetry, we need to deal only with
	\begin{equation*}
		W_3(1,2,3) = \nabla^\perp w_R(x_1 - x_2)\cdot\nabla^\perp w_R(x_1 - x_3).
	\end{equation*}
	Using again the Fourier transform, we have
	\begin{equation*}
		W_3(1,2,3) = \int\d{}k\chi_R(k)\dfrac{e_{k^\perp}}{\vert k\vert}e^{\mathbf{i}k\cdot(x_1 - x_2)}\int\d{}k'\chi_R(k')\dfrac{e_{k'^\perp}}{\vert k'\vert}e^{\mathbf{i}k'\cdot(x_1 - x_3)}.
	\end{equation*}
	Developing the complex exponentials and using the oddness of $k \mapsto \frac{\hat{\chi}_R(k)}{\vert k\vert}e_{k^\perp}$ yields
	\begin{multline}
		\label{eq:three_body_potential_fourier}
		W_3(1,2,3) =\iint\d{}k\d{}k'\hat{\chi}_R(k)\hat{\chi}_R(k')\dfrac{e_{k^\perp}}{\vert k\vert}\cdot \dfrac{e_{k'^\perp}}{\vert k'\vert}\Big(\cos(k\cdot x_1)\cos(k'\cdot x_1) \sin(k\cdot x_2)\sin(k'\cdot x_3)\\
		\begin{aligned}[b]
			&+ \sin(k\cdot x_1)\sin(k'\cdot x_1) \cos(k\cdot x_2)\cos(k'\cdot x_3)\\
			&- \cos(k\cdot x_1)\sin(k'\cdot x_1) \sin(k\cdot x_2)\cos(k'\cdot x_3)\\
			&- \sin(k\cdot x_1)\cos(k'\cdot x_1) \cos(k\cdot x_2)\sin(k'\cdot x_3)\Big).
		\end{aligned}
	\end{multline}
	Using once more Theorem~\ref{th:de_finetti} and Proposition~\ref{prop:plane_wave_estimate}, we find
	\begin{equation}
		\label{eq:remainder_de_finetti_W3}
		\Tr\big(W_3\Delta\gamma_{i,K}^{(3)}\big) \geq -C\sqrt{\dfrac{\log R^{-1}}{K}}N^{2i\varepsilon}.
	\end{equation}
	Here we again split the integral over $k$ between $\vert k\vert \leq N^{i\varepsilon}$ and $\vert k\vert > N^{i\varepsilon},$ and we did the same for $k'$.
	\\
	
	Inserting \eqref{eq:remainder_de_finetti_kinetic}--\eqref{eq:remainder_de_finetti_W3} into \eqref{eq:hamiltonian_lower_bound_before_de_finetti}, and using that $\mu_{i,K}$ is a probability measure, we obtain
	\begin{equation}
		\label{eq:hamiltonian_3B_lower_bound}
		\Tr\big(\tilde{H}_3^R\gamma_{i,K}^{(3)}\big) \geq \int_{\mathcal{S}(\bbP_i\mathfrak{H})}\Tr\big(\tilde{H}_3^R\gamma^{\otimes 3}\big)\d{}\mu_{i,K}(\gamma) - C\sqrt{\dfrac{\log R^{-1}}{K}}N^{2i\varepsilon}.
	\end{equation}
	As a result of \eqref{eq:hamiltonian_lower_bound_proj_decomposition_high_occupancy} and \eqref{eq:hamiltonian_3B_lower_bound}, we have
	\begin{align*}
		N^{-1}\left\langle\Psi,H_N^R\Psi\right\rangle &\geq \sum_{i = 1}^M\sum_K{N \choose 3}^{-1}{K \choose 3}\Tr\Gamma_{i,K}\int_{\mathcal{S}(\bbP_i\mathfrak{H})}\Tr\big(\tilde{H}_3^R\gamma^{\otimes 3}\big)\d{}\mu_{i,K}(\gamma)\\
		&\phantom{\geq} - C\sqrt{\dfrac{\log R^{-1}}{K}}\sum_{i = 1}^M\sum_K{N \choose 3}^{-1}{K \choose 3}N^{2i\varepsilon}\Tr\Gamma_{i,K}\\
		&\phantom{\geq} - C(N^{-\varepsilon} + N^{-(1 - \kappa)/2 + 2\varepsilon})\Tr\big((C + h)\Gamma^{(1)}\big). \numberthis \label{eq:hamiltonian_lower_bound_after_de_finetti}
	\end{align*}
	On the one hand, using ${N \choose 3}^{-1}{K \choose 3} \leq 1$ and that $K \geq N - MN^{1-\delta\varepsilon}$, we can bound the second term in \eqref{eq:hamiltonian_lower_bound_after_de_finetti} by
	\begin{equation*}
		\sqrt{\dfrac{\log R^{-1}}{K}}\sum_{i = 1}^M\sum_K{N \choose 3}^{-1}{K \choose 3}N^{2i\varepsilon}\Tr\Gamma_{i,K} \leq CN^{\delta\varepsilon/2 - 1/2}\sqrt{\log R^{-1}}\sum_{\underline{J}}N^{2\imax(\underline{J})\varepsilon}\Tr\Gamma_{\underline{J}},
	\end{equation*}
	where we recall that $\imax$ was defined in \eqref{eq:imax_def_poly}. But, thanks to $1 \leq N^{\delta\varepsilon}j_{i_{\textmd{max}}}/N$ and Lemma \ref{lemma:state_three_projections_full_trace}, we have
	\begin{equation*}
		N^{2\imax\varepsilon}\Tr\Gamma_{\underline{J}} \leq N^{2\varepsilon}\Tr\big(P_{\imax(\underline{J})}(C + h)P_{\imax(\underline{J})}\Gamma_{\underline{J}}^{(1)}\big) \leq N^{2\varepsilon}\Tr\big((C + h)\Gamma^{(1)}_{\underline{J}}\big).
	\end{equation*}
	Hence, by applying Lemma \ref{lemma:occupancy_wavefunction_decomposition}, we deduce
	\begin{equation*}
		\sqrt{\dfrac{\log R^{-1}}{K}}\sum_{i = 1}^M\sum_K{N \choose 3}^{-1}{K \choose 3}N^{2i\varepsilon}\Tr\Gamma_{i,K} \leq CN^{(3\delta/2 + 2)\varepsilon - 1/2}\sqrt{\log R^{-1}}\Tr\big((C + h)\Gamma^{(1)}\big).
	\end{equation*}
	This means that this term is much smaller than the kinetic energy under the condition
	\begin{equation}
		\label{eq:relation_delta_varepsilon}
		\left(3\delta/2 + 2\right)\varepsilon < (1 - \kappa)/2
	\end{equation}
	(with $\kappa = 0$ in the polynomial case). On the other hand, to simplify the first term in \eqref{eq:hamiltonian_lower_bound_after_de_finetti}, we define the Borel measure
	\begin{equation*}
		\label{eq:borel_measure_de_finetti}
		\mu_N = \sum_{i = 1}^M\sum_K{N \choose 3}^{-1}{K \choose 3}\Tr(\Gamma_{i,K})\mu_{i,K}
	\end{equation*}
	and we extend the definition \eqref{eq:average_field_energy_functional} of the average-field functional $\cE_R^\af$ to mixed-states $\gamma$ through
	\begin{equation*}
		\cE_R^\af[\gamma] = \Tr\big(\tilde{H}_3^R\gamma^{\otimes 3}\big).
	\end{equation*}
	
	Summing up, the inequality \eqref{eq:hamiltonian_lower_bound_after_de_finetti} simplifies to
	\begin{equation*}
		N^{-1}\langle\Psi,H_N^R\Psi\rangle \geq \int_{\cS(\bbP_M\mathfrak{H})}\cE_R^\af[\gamma]\d{}\mu_N(\gamma)
		- C(N^{-\varepsilon} + N^{(3\delta/2 + 2)\varepsilon - (1 - \kappa)/2}\sqrt{\log N})\Tr\big((C + h)\Gamma^{(1)}\big),
	\end{equation*}
	with $\kappa = 0$ in the polynomial case \eqref{eq:smearing_radius_polynomial}. Moreover, the same inequality is true with the left-hand side replaced by $\Tr(\tilde{H}_3^R\Gamma^{(3)})$. Thus, injecting this into \eqref{eq:kinetic_energy_bounded_a_priori_intermediate} and using the bound
	\begin{equation*}
		\cE_R^\af[\gamma] \geq \Tr V\gamma \geq -C,
	\end{equation*}
	we deduce \eqref{eq:kinetic_energy_bounded_a_priori}.\footnote{Recall that, at this stage, this bound had not yet been proved for $R = e^{-N^\kappa}$ with $1/2\leq \kappa < 1$.} The previous estimate is shown using the spectral decomposition of $\gamma$; we refer to \cite[Section 3.6]{Girardot2020averageFA} for more detail.
	
	To conclude the proof of Theorem~\ref{th:convergence_gse_average_field}, we wish to use \eqref{eq:kinetic_energy_bounded_a_priori} and
	\begin{equation}
		\label{eq:average_field_functional_liminf}
		\liminf_{N\rightarrow\infty}\int_{\cS(\bbP_M\mathfrak{H})}\cE_R^\af[\gamma]\d{}\mu_N(\gamma) \geq e^\af.
	\end{equation}
	To prove the latter, we need to show that the sequence $(\mu_N)_N$ converges to a probability measure supported on pure states $\vert u\rangle\langle u\vert$. For this, we show that the measure $\mu_N$ satisfies
	\begin{equation}
		\label{eq:borel_measure_de_finetti_almost_proba}
		1 - CN^{-2\varepsilon} \leq \int_{\cS(\bbP_M\mathfrak{H})}\mu_N(\gamma) \leq 1.
	\end{equation}
	Granted that \eqref{eq:borel_measure_de_finetti_almost_proba} is true, we can follow \cite[Section 3.7.]{Girardot2020averageFA} (see references therein) to prove that $(\mu_N)_N$ indeed converges to a probability measure supported on pure states $\vert u\rangle\langle u\vert$, from which \eqref{eq:average_field_functional_liminf} and thus \eqref{eq:energy_converence_main_result} follow. The upper bound in \eqref{eq:borel_measure_de_finetti_almost_proba} is a consequence of the estimate ${N \choose 3}^{-1}{K \choose 3} \leq 1$ and the identity
	\begin{equation*}
		\sum_{i = 0}^{M + 1}\sum_K\Tr\Gamma_{i,K} = 1.
	\end{equation*}
	For the lower bound, we use
	\begin{align*}
		\int_{\cS(\bbP_M\mathfrak{H})}\mu_N(\gamma) &= \sum_{i = 1}^M\sum_K{N \choose 3}^{-1}{K \choose 3}\Tr\Gamma_{i,K} = \sum_{i = 1}^M\sum_K\Tr\big(\bbP_i^{\otimes 3}\Gamma_{i,K}^{(3)}\big)\\
		&= \sum_{i = 0}^M\sum_K\Tr\Gamma_{i,K} - \sum_{K}\Tr\Gamma_{0,K}\\
		&\phantom{=} -\sum_{i = 1}^M\sum_K\big(\Tr\big(\bbP_i^{\otimes 2}\otimes_{\textmd{s}}\bbQ_i\Gamma_{i,K}^{(3)}\big) + \Tr\big(\bbP_i\otimes_{\textmd{s}}\bbQ_i^{\otimes 2}\Gamma_{i,K}^{(3)}\big) + \Tr\big(\bbQ_i^{\otimes 3}\Gamma_{i,K}^{(3)}\big)\big)\\
		&\geq 1 - C(N^{-2\varepsilon} + N^{-2M\varepsilon})\Tr\big((C + h)\Gamma^{(1)}\big)
	\end{align*}
	and \eqref{eq:kinetic_energy_bounded_a_priori}. To bound the error terms, we used
	\begin{equation*}
		\sum_{i = 1}^M\sum_K\Tr\big(\bbQ_i\Gamma_{i,K}^{(1)}\big) \leq CN^{-2\varepsilon}\sum_{i = 1}^M\sum_{K}\Tr\big(\bbQ_i(C + h)\bbQ_i\Gamma_{i,K}^{(1)}\big) \leq CN^{-2\varepsilon}\Tr\big((C + h)\Gamma^{(1)}\big),
	\end{equation*}
	and
	\begin{align*}
		\sum_K\Tr\Gamma_{0,K} \leq CN^{-2M\varepsilon}\sum_K\dfrac{N}{K}\Tr\big(P_{M + 1}hP_{M + 1}\Gamma_{0,K}^{(1)}\big) &\leq CN^{-2M\varepsilon}\sum_K\Tr\big(P_{M + 1}hP_{M + 1}\Gamma_{0,K}^{(1)}\big)\\
		&\leq CN^{-2M\varepsilon}\Tr\big(h\Gamma^{(1)}\big).
	\end{align*}
	The last estimate results from the fact that we are summing over $N - MN^{1 - \delta\varepsilon} \leq K \leq N$, which follows from the convention that $\imax = 0$ (corresponding to $i = 0$) when $j_{k} \leq N^{1 - \delta\varepsilon}$ for all $k \in\{1,\dots,M\}$. This concludes the proof of Theorem \ref{th:convergence_gse_average_field}.
	
	\section{Proof of the main estimates}
	
	\label{section:proof_main_estimates}
	
	\subsection{High-momenta estimates: proof of Lemma~\ref{lemma:high_momenta_estimates}}
	
	We begin with the proof of the estimate \eqref{eq:high_momenta_estimate_W3}. Thanks to the Cauchy--Schwarz inequality and the estimates \eqref{eq:smeared_coulomb_potential_estimates}~and~\eqref{eq:three-body_term_bound_nonneg}, we have
	\begin{multline*}
		\left\vert\Tr\left(P_{M+1} \otimes Q_1 \otimes Q_2W_3Q_3\otimes Q_4\otimes Q_5\Gamma\right)\right\vert\\
		\begin{aligned}[t]
			&\leq
			\begin{multlined}[t]
				CN^{2\varepsilon}R^{-2}\Tr\big(P_{M+1}\otimes Q_1 \otimes Q_2\Gamma\big)\\
				+ CN^{-2\varepsilon}\Tr\big(Q_3\otimes Q_4\otimes Q_5\big(1 - \Delta_1 - \Delta_2 - \Delta_3\big)Q_3\otimes Q_4\otimes Q_5\Gamma\big)
			\end{multlined}\\
			&\leq CN^{-2\varepsilon}\Tr\big((C + h)\Gamma^{(1)}\big).
		\end{aligned}
	\end{multline*}
	In the second inequality we used
	\begin{equation*}
		N^{2\varepsilon}R^{-2} \leq N^{-2\varepsilon}N^{2M\varepsilon} \quad \textmd{and} \quad P_{M + 1}N^{2M\varepsilon}P_{M + 1}\leq P_{M + 1}hP_{M + 1}.
	\end{equation*}
	The former follows from the definition of $R$ and assumption \eqref{eq:relation_M_epsilon}. The latter is an immediate consequence of the definition of $P_{M + 1}$.
	
	We now prove \eqref{eq:high_momenta_estimate_W2}. We only take care of
	\begin{equation*}
		W_2(1,2) = \p_1\cdot\nabla^\perp w_R(x_1 - x_2)
	\end{equation*}
	as the other terms are analogous. Due to $\nabla_1\cdot \nabla^\perp w_R(x_1 - x_2) = 0$, we have
	\begin{equation*}
		\mathbf{p}_1\cdot\nabla^\perp w_R(x_1 - x_2) = \nabla^\perp w_R(x_1 - x_2)\cdot\mathbf{p}_1.
	\end{equation*}
	Combining this with the Cauchy--Schwarz inequality and the estimate \eqref{eq:smeared_coulomb_potential_estimates}, we find
	\begin{align*}
		\left\vert\Tr\left(P_{M+1}\otimes Q_1\otimes Q_2W_2(1,2)Q_3\otimes Q_4\otimes Q_5\Gamma\right)\right\vert &\leq 
		\begin{multlined}[t]
			CR^{-2}N^{2\varepsilon}\Tr\left(P_{M+1}\otimes Q_1\otimes Q_2\Gamma\right)\\
			+ CN^{-2\varepsilon}\Tr\big(Q_3\otimes Q_4\otimes Q_5(-\Delta_1)Q_3\otimes Q_4\otimes Q_5\Gamma\big)
		\end{multlined}\\
		&\leq CN^{-2\varepsilon}\Tr\big((C + h)\Gamma^{(1)}\big).
	\end{align*}
	This concludes the proof of Lemma~\ref{lemma:high_momenta_estimates}.
	
	\subsection{Low-occupancy estimates: Proof of Lemma~\ref{lemma:low_occupancy_estimates}}
	
	In this section we provide the proof of Lemma~\ref{lemma:low_occupancy_estimates}. Although the proof used in the exponential case covers polynomial scalings as well, we first present a simpler proof for the polynomial case for the reader's convenience. Both proofs rely on the following lemma, for which we define
	\begin{equation*}
		\bbP_{jk} = \bbP_j\otimes \bbP_k \quad \textmd{and} \quad \bbP_{jk\ell} = \bbP_j\otimes \bbP_k\otimes \bbP_\ell,
	\end{equation*}
	with $\bbP_j = \mathds{1}(h \leq N^{2j\varepsilon})$.
	
	\begin{lemma}
		\label{lemma:plane_wave_estimates_W2_W3}
		Let $W_2(1,2)$ and $W_3(1,2,3)$ be as in \eqref{eq:two_and_three_body_term_def} and \eqref{eq:two_and_three_body_term_def}, with projections as in \eqref{eq:projections_momentum_space}. Then, there exists a universal constant $C > 0$ such that, for any $i_1,i_2,i_1',i_2'$ and $f,g\in\mathfrak{H}^2$,
		\begin{equation}
			\label{eq:plane_wave_estimate_W2}
			\langle \bbP_{i_1i_2}f,W_2(1,2)\bbP_{i_1'i_2'}g\rangle \leq CN^{\max(i_2,i_2')\varepsilon}\Vert \bbP_{i_1i_2}f\Vert\Vert \p_1 \bbP_{i_1'i_2'}g\Vert.
		\end{equation}
		and
		\begin{equation}
			\label{eq:plane_wave_estimate_W2_second}
			\langle \bbP_{i_1i_2}f,W_2(1,2)\bbP_{i_1'i_2'}g\rangle \leq CN^{(i_1 + i_1')\varepsilon}\Vert \bbP_{i_1i_2}f\Vert\Vert\bbP_{i_1'i_2'}g\Vert.
		\end{equation}
		Moreover, for any $i_1,i_2,i_3$,
		\begin{equation}
			\label{eq:plane_wave_estimate_W3}
			P_{i_1i_2i_3}W_3(1,2,3)P_{i_1i_2i_3} \leq CN^{(i_2 + i_3)\varepsilon}.
		\end{equation}
	\end{lemma}
	
	\begin{proof}[Proof of Lemma~\ref{lemma:plane_wave_estimates_W2_W3}]
		We start with \eqref{eq:plane_wave_estimate_W3}. Define
		\begin{equation*}
			\chi_R = \dfrac{\mathds{1}_{B(0,R)}}{\pi R^2}
		\end{equation*}
		and write
		\begin{align*}
			W_3(1,2,3) = \iint\d{}k\d{}k'\hat{\chi}_R(k)\hat{\chi}_R(k')\dfrac{\e_{k^\perp}}{\vert k\vert}\cdot\dfrac{e_{k'^\perp}}{\vert k'\vert}\big[&\cos(k\cdot x_1)\cos(k'\cdot x_1)\sin(k\cdot x_2)\sin(k'\cdot x_3)\\
			&+ \sin(k\cdot x_1)\sin(k'\cdot x_1)\cos(k\cdot x_2)\cos(k'\cdot x_3)\\
			&- \cos(k\cdot x_1)\sin(k'\cdot x_1)\sin(k\cdot x_2)\cos(k'\cdot x_3)\\
			&- \sin(k\cdot x_1)\cos(k'\cdot x_1)\cos(k\cdot x_2)\sin(k'\cdot x_3)\big],
		\end{align*}
		as in the decomposition \eqref{eq:three_body_potential_fourier}. Then, we split the integral over $\vert k \vert \leq N^{i_2\varepsilon}$ and $\vert k \vert > N^{i_2\varepsilon}$, and likewise for $k'$. Conjugating with $P_{i_1i_2i_3}$ and applying Proposition~\ref{prop:plane_wave_estimate} to $P_{i_2}\sin(k\cdot x_2)P_{i_2}$ and $P_{i_3}\sin(k'\cdot x_3)P_{i_3}$, we obtain
		\begin{multline*}
			P_{i_1i_2i_3}W_3(1,2,3)P_{i_1i_2i_3}\\
			\begin{aligned}[t]
				&\leq C\Big(\int_{\vert k\vert \leq N^{i_2\varepsilon}}\dfrac{\d{}k}{\vert k\vert} + \int_{\vert k\vert > N^{i_2\varepsilon}}\d{}k\dfrac{N^{2i_2\varepsilon}}{\vert k\vert^3}\Big)\Big(\int_{\vert k'\vert \leq N^{i_3\varepsilon}}\dfrac{\d{}k'}{\vert k'\vert} + \int_{\vert k'\vert > N^{i_3\varepsilon}}\d{}k'\dfrac{N^{2i_3\varepsilon}}{\vert k'\vert^3}\Big)\\
				&\leq CN^{(i_2 + i_3)\varepsilon}.
			\end{aligned}
		\end{multline*}
		Here, we used $\Vert\hat{\chi}_R\Vert_{L^\infty} \leq \Vert\chi_R\Vert_{L^1}\leq C$.
		
		Now, we prove \eqref{eq:plane_wave_estimate_W2_second} for $f$ and $g$ smooth with compact support; the result follows from a density argument. We again write in Fourier
		\begin{equation*}
			W_2(1,2) = 4\pi\int\d{}k\dfrac{\hat{\chi}_R(k)}{\vert k\vert}\left(\sin(k\cdot x_1)\cos(k\cdot x_2) - \cos(k\cdot x_1)\sin(k\cdot x_2)\right)e_{k^\perp}\cdot\mathbf{p}_1,
		\end{equation*}
		as in \eqref{eq:two_body_mixed_potential_fourier}. We suppose without loss of generality that $i_1 \geq i_1'$ and we split the integral over $\vert k\vert \leq N^{i_1\varepsilon}$ and $\vert k\vert > N^{i_1\varepsilon}$. On the first sector, we have
		\begin{multline*}
			\left\vert\int_{\vert k\vert \leq N^{i_1\varepsilon}}\d{}k\dfrac{\hat{\chi}_R(k)}{\vert k\vert}\left\langle \bbP_{i_1i_2}f\left(\sin(k\cdot x_1)\cos(k\cdot x_2) - \cos(k\cdot x_1)\sin(k\cdot x_2)\right)e_{k^\perp}\cdot\mathbf{p}_1 \bbP_{i_1'i_2'}g\right\rangle\right\vert\\
			\leq CN^{i_1\varepsilon}\Vert \bbP_{i_1i_2}f\Vert \Vert \p_1 \bbP_{i_1'i_2'}g\Vert \leq CN^{(i_1 + i_1')\varepsilon}\bbP_{i_1i_2}f\Vert \Vert\bbP_{i_1'i_2'}g\Vert.
		\end{multline*}
		On the second sector, we write $\sin(k\cdot x_1) = -k\cdot\nabla\cos(k\cdot x_1)/\vert k\vert^2$ (resp. $\cos(k\cdot x_1) = k\cdot\nabla\sin(k\cdot x_1)/\vert k\vert^2$) and use two successive integrations by parts on $x_1$ to obtain
		\begin{multline*}
			\left\vert\int_{\vert k\vert > N^{i_1\varepsilon}}\d{}k\dfrac{\hat{\chi}_R(k)}{\vert k\vert}\left\langle \bbP_{i_1i_2}f\left(\sin(k\cdot x_1)\cos(k\cdot x_2) - \cos(k\cdot x_1)\sin(k\cdot x_2)\right)e_{k^\perp}\cdot\mathbf{p}_1 \bbP_{i_1'i_2'}g\right\rangle\right\vert\\
			\leq CN^{(i_1 + i_1')\varepsilon}\Vert \bbP_{i_1i_2}f\Vert \Vert\bbP_{i_1'i_2'}g\Vert.
		\end{multline*}
		Here, we used the bound \eqref{eq:laplacian_and_square_bdd_projections}.
		
		The bound \eqref{eq:plane_wave_estimate_W2} is proved similarly: we split between $\vert k\vert \leq N^{\max(i_2,i_2')\varepsilon}$ and $\vert k\vert > N^{\max(i_2,i_2')\varepsilon}$, and we use two integrations by parts on the variable $x_2$ in the second sector. We omit the details.
	\end{proof}

	\begin{proof}[Proof of Lemma~\ref{lemma:low_occupancy_estimates} in the polynomial case]
		We begin by proving \eqref{eq:low_momenta_estimate_W3} since it is slightly easier. We provide the details only for
		\begin{equation*}
			W_3(1,2,3) = \nabla^\perp w_R(x_1 - x_2)\cdot\nabla^\perp w_R(x_1 - x_3)
		\end{equation*}
		as the other terms are dealt with analogously. Using the Cauchy--Schwarz inequality and Lemmas~\ref{lemma:state_three_projections_full_trace}~and~\ref{lemma:plane_wave_estimates_W2_W3}, we expand, for all $\tau > 0$,
		\begin{multline}
			\big\vert\Tr\big(P_{i_1i_2i_3}W_3(1,2,3)P_{i_1'i_2'i_3'}\Gamma_{\underline{J}^{(i_1i_2i_3;i_1'i_2'i_3')}}^{(3)}\big)\big\vert \leq \tau CN^{(i_2 + i_3)\varepsilon}\Tr\big(P_{i_1i_2i_3}\Gamma_{\underline{J}^{(i_1i_2i_3)}}^{(3)}\big)\\
			+ \tau^{-1}CN^{(i_2' + i_3')\varepsilon}\Tr\big(P_{i_1'i_2'i_3'}\Gamma_{\underline{J}^{(i_1'i_2'i_3')}}^{(3)}\big). \label{eq:low_momenta_estimate_poly_W3_proof}
		\end{multline}
		On the one hand thanks to Lemma \ref{lemma:state_three_projections_full_trace}, the first term can be bounded by
		\begin{equation*}
			\Tr\big(P_{i_1i_2i_3}\Gamma_{\underline{J}^{(i_1i_2i_3)}}^{(3)}\big) \leq N^{-3}(j_{i_1} + 1)(j_{i_2} + 1)(j_{i_3} + 1)\Tr\Gamma_{\underline{J}^{(i_1i_2i_3)}}.
		\end{equation*}
		Thus, using $j_{i_1} \leq N^{1 - \delta\varepsilon}$, the estimate
		\begin{equation}
			\label{eq:kinetic_part_lower_bound_N}
			P_iN^{2(i - 1)\varepsilon}P_i \leq CP_i(C + h)P_i
		\end{equation}
		and Lemma~\ref{lemma:state_three_projections_full_trace}, we deduce
		\begin{equation*}
			N^{(i_2 + i_3)\varepsilon}\Tr\big(P_{i_1i_2i_3}\Gamma_{\underline{J}^{(i_1i_2i_3)}}^{(3)}\big) \leq CN^{-\delta\varepsilon + 2\varepsilon}\Tr\big((C + h)\Gamma_{\underline{J}^{(i_1i_2i_3)}}^{(1)}\big).
		\end{equation*}
		On the other hand, thanks to \eqref{eq:kinetic_part_lower_bound_N}, we have
		\begin{equation*}
			N^{(i_2' + i_3')\varepsilon}\Tr\big(P_{i_1'i_2'i_3'}\Gamma_{\underline{J}^{(i_1'i_2'i_3')}}^{(3)}\big) \leq N^{2\varepsilon}\Tr\big((C + h)\Gamma_{\underline{J}^{(i_1'i_2'i_3')}}^{(1)}\big).
		\end{equation*}
		Injecting the previous two inequalities into \eqref{eq:low_momenta_estimate_poly_W3_proof} and optimising over $\tau$, we find
		\begin{multline*}
			\big\vert\Tr\big(P_{i_1i_2i_3}W_3(1,2,3)P_{i_1'i_2'i_3'}\Gamma_{\underline{J}^{(i_1i_2i_3;i_1'i_2'i_3')}}^{(3)}\big)\big\vert \leq CN^{-(\delta/2 - 2)\varepsilon}\Tr\big((C + h)\Gamma_{\underline{J}^{(i_1i_2i_3)}}^{(1)}\big)\\
			+ CN^{-(\delta/2 - 2)\varepsilon}\Tr\big((C + h)\Gamma_{\underline{J}^{(i_1'i_2'i_3')}}^{(1)}\big).
		\end{multline*}
		Summing over $\underline{J}$, this becomes
		\begin{equation*}
			\sum_{\underline{J}}\big\vert\Tr\big(P_{i_1i_2i_3}W_3(1,2,3)P_{i_1'i_2'i_3'}\Gamma_{\underline{J}^{(i_1i_2i_3;i_1'i_2'i_3')}}^{(3)}\big)\big\vert \leq CN^{-(\delta/2 - 2)\varepsilon}\Tr\big((C + h)\Gamma^{(1)}\big).
		\end{equation*}
		Finally, we sum over $i_1,\dots,i_3'$ and use that $M$ is a constant to obtain \eqref{eq:low_momenta_estimate_W3}.
		
		We now prove \eqref{eq:low_momenta_estimate_W2}. Again, we bound only
		\begin{equation*}
			W_2(1,2) = \p_1\cdot w_R(x_1 - x_2) + w_R(x_1 - x_2)\cdot\p_1
		\end{equation*}
		and leave the rest to the reader. Thanks to the Cauchy--Schwarz inequality and Lemmas~\ref{lemma:state_three_projections_full_trace}~and~\ref{lemma:plane_wave_estimates_W2_W3}, we have
		\begin{multline}
			\label{eq:low_occupancy_W2_bound_poly}
			\big\vert\Tr\big(P_{i_1i_2i_3}W_2(1,2)P_{i_1'i_2'i_3'}\Gamma_{\underline{J}^{(i_1i_2i_3;i_1'i_2'i_3')}}^{(3)}\big)\big\vert\\
			\begin{aligned}[b]
				&\leq CN^{\max(i_2,i_2')\varepsilon}\tau\Tr\big(P_{i_1i_2i_3}\Gamma_{\underline{J}^{(i_1i_2i_3)}}^{(3)}\big)\\
				&\phantom{\leq} + CN^{\max(i_2,i_2')\varepsilon}\tau^{-1}\Tr\big(P_{i_1'i_2'i_3'}(-\Delta_1)P_{i_1'i_2'i_3'}\Gamma_{\underline{J}^{(i_1'i_2'i_3')}}^{(3)}\big)\\
				&\leq CN^{\max(i_2,i_2')\varepsilon}N^{-3}\tau (j_{i_1} + 1)(j_{i_2} + 1)(j_{i_3} + 1)\Tr\Gamma_{\underline{J}^{(i_1i_2i_3)}}\\
				&\phantom{\leq} + CN^{\max(i_2,i_2')\varepsilon}N^{-3}\tau^{-1}N^{2i_1'\varepsilon}(j_{i_1'} + 1)(j_{i_2'} + 1)(j_{i_3'} + 1)\Tr\Gamma_{\underline{J}^{(i_1'i_2'i_3')}},
			\end{aligned}
		\end{multline}
		for all $\tau > 0$. On the one hand, when $i_2 \geq i_2'$, we use $j_{i_1} \leq N^{1 - \delta\varepsilon}$, the estimate \eqref{eq:kinetic_part_lower_bound_N}, Lemma~\ref{lemma:state_three_projections_full_trace}, and optimise over $\tau$ to show that
		\begin{align*}
			\big\vert\Tr\big(P_{i_1i_2i_3}W_2(1,2)P_{i_1'i_2'i_3'}\Gamma_{\underline{J}^{(i_1i_2i_3;i_1'i_2'i_3')}}^{(3)}\big)\big\vert &\leq CN^{-(\delta/2  - 2)\varepsilon}\Tr\big(P_{i_2i_3}(C + h_1)P_{i_2i_3}\Gamma_{\underline{J}^{(i_1i_2i_3)}}^{(2)}\big)\\
			&\phantom{\leq} + CN^{-(\delta/2  - 2)\varepsilon}\Tr\big(P_{i_1'i_2'i_3'}(C + h_1)P_{i_1'i_2'i_3'}\Gamma_{\underline{J}^{(i_1'i_2'i_3')}}^{(3)}\big)\\
			&\leq CN^{-(\delta/2  - 2)\varepsilon}\Tr\big((C + h)\Gamma_{\underline{J}^{(i_1i_2i_3)}}^{(1)}\big)\\
			&\phantom{\leq} + CN^{-(\delta/2  - 2)\varepsilon}\Tr\big((C + h)\Gamma_{\underline{J}^{(i_1'i_2'i_3')}}^{(1)}\big).
		\end{align*}
		On the other hand, when $i_2 < i_2'$, an appropriate choice of $\tau$ in \eqref{eq:low_occupancy_W2_bound_poly} yields
		\begin{align*}
			\big\vert\Tr\big(P_{i_1i_2i_3}W_2(1,2)P_{i_1'i_2'i_3'}\Gamma_{\underline{J}^{(i_1i_2i_3;i_1'i_2'i_3')}}^{(3)}\big)\big\vert &\leq CN^{-(\delta/2 - 2)\varepsilon}\Tr\big(P_{i_2'i_3}(C + h_1)P_{i_2'i_3}\Gamma_{\underline{J}^{(i_1i_2i_3)}}^{(2)}\big)\\
			&\phantom{\leq} + CN^{-(\delta/2 - 2)\varepsilon}\Tr\big(P_{i_1'i_2i_3'}(C + h_1)P_{i_1'i_2i_3'}\Gamma_{\underline{J}^{(i_1'i_2'i_3')}}^{(3)}\big)\\
			&\leq CN^{-(\delta/2 - 2)\varepsilon}\Tr\big((C + h)\Gamma_{\underline{J}^{(i_1i_2i_3)}}^{(1)}\big)\\
			&\phantom{\leq} + CN^{-(\delta/2 - 2)\varepsilon}\Tr\big((C + h)\Gamma_{\underline{J}^{(i_1'i_2'i_3')}}^{(1)}\big).
		\end{align*}
		Gathering the previous estimates and carrying out all the sums gives \eqref{eq:low_momenta_estimate_W2}. This concludes the proof of Lemma~\ref{lemma:low_occupancy_estimates} for polynomially decreasing radii.
	\end{proof}

	\begin{proof}[Proof of Lemma~\ref{lemma:low_occupancy_estimates} in the exponential case]
		\textit{Analysis of $W_3$.} We first prove \eqref{eq:low_momenta_estimate_W3}. We give the proof in full detail for
		\begin{equation*}
			W_3(1,2,3) = \nabla^\perp w_R(x_1 - x_2)\cdot \nabla^\perp w_R(x_1 - x_3),
		\end{equation*}
		and we assume that $i_1 \geq i_2 \geq i_3$, as well as $i_1' \geq i_2' \geq i_3'$ and $i_1 \geq i_1'$. The rest is similar and left to the reader. To prove \eqref{eq:low_momenta_estimate_W3}, we make a distinction between $\vert i_1 - i_3\vert \leq \Lambda$ and $\vert i_1 - i_3\vert > \Lambda$, and likewise for $(i_1',i_3')$ and $(i_1,i_1')$. Since we only need to use $j_{i_1} \leq N^{1 - \delta\varepsilon}$ when $\vert i_1 - i_3\vert,\vert i_1' - i_3'\vert,\vert i_1 - i_1'\vert \leq \Lambda$, we begin by treating the other cases. Namely, we bound first
		\begin{equation}
			\label{eq:low_momenta_estimate_expo_W3_far_first}
			\sum_{\underline{J}}\sum_{\substack{i_1 \geq i_2 \geq i_3\\ i_1 > i_3 + \Lambda}}\sum_{\substack{i_1' \geq i_2' \geq i_3'}}\big\vert\Tr\big(P_{i_1i_2i_3}W_3(1,2,3)P_{i_1'i_2'i_3'}\Gamma_{\underline{J}^{(i_1i_2i_3;i_1'i_2'i_3')}}^{(3)}\big)\big\vert
		\end{equation}
		and
		\begin{equation}
			\label{eq:low_momenta_estimate_expo_W3_far_second}
			\sum_{\underline{J}}\sum_{\substack{i_1 \geq i_2 \geq i_3\\ \vert i_1 - i_3\vert \leq \Lambda}}\sum_{\substack{i_1' \geq i_2' \geq i_3'\\ \vert i_1' - i_3'\vert \leq \Lambda\\ i_1 > i_1' + \Lambda}}\big\vert\Tr\big(P_{i_1i_2i_3}W_3(1,2,3)P_{i_1'i_2'i_3'}\Gamma_{\underline{J}^{(i_1i_2i_3;i_1'i_2'i_3')}}^{(3)}\big)\big\vert.
		\end{equation}
		
		In order to bound \eqref{eq:low_momenta_estimate_expo_W3_far_first}, we use the Cauchy--Schwarz inequality and Lemmas~\ref{lemma:state_three_projections_full_trace}~and~\ref{lemma:plane_wave_estimates_W2_W3} to write
		\begin{multline*}
			\big\vert \Tr\big(P_{i_1i_2i_3}W_3(1,2,3)P_{i_1'i_2'i_3'}\Gamma_{\underline{J}^{(i_1i_2i_3;i_1'i_2'i_3')}}^{(3)}\big)\big\vert\\
			\leq \tau CN^{(i_2 + i_3)\varepsilon}N^{-3}(j_{i_1} + 1)(j_{i_2} + 1)(j_{i_3} + 1)\Tr\Gamma_{\underline{J}^{(i_1i_2i_3)}}\\
			+ \tau^{-1}CN^{(i_2' + i_3')\varepsilon}N^{-3}(j_{i_1'} + 1)(j_{i_2'} + 1)(j_{i_3'} + 1)\Tr\Gamma_{\underline{J}^{(i_1'i_2'i_3')}},
		\end{multline*}
		for all $\tau > 0$. Swapping $j_{i_2}$ and $j_{i_2'}$ by an appropriate choice of $\tau$, using the estimate
		\begin{equation}
			\label{eq:kinetic_part_lower_bound_N_2}
			P_iN^{2(i - 1)\varepsilon}P_i \leq CP_i(C + h)P_i,
		\end{equation}
		and Lemma~\ref{lemma:state_three_projections_full_trace}, we thus obtain
		\begin{multline*}
			\big\vert \Tr\big(P_{i_1i_2i_3}W_3(1,2,3)P_{i_1'i_2'i_3'}\Gamma_{\underline{J}^{(i_1i_2i_3;i_1'i_2'i_3')}}^{(3)}\big)\big\vert\\
			\begin{aligned}[t]
				&\leq CN^{(i_2 - i_1)\varepsilon/2}N^{(i_3 - i_1 + 4)\varepsilon/2}N^{(i_2' - i_1')\varepsilon/2}N^{(i_3' - i_1')\varepsilon/2}\Tr\big(P_{i_1i_2'i_3}(C + h_1)P_{i_1i_2'i_3}\Gamma_{\underline{J}^{(i_1i_2i_3)}}^{(3)}\big)\\
				&\phantom{\leq} + CN^{(i_2 - i_1)\varepsilon/2}N^{(i_3 - i_1 + 4)\varepsilon/2}N^{(i_2' - i_1')\varepsilon/2}N^{(i_3' - i_1')\varepsilon/2}\Tr\big(P_{i_1'i_2i_3'}(C + h_1)P_{i_1'i_2i_3'}\Gamma_{\underline{J}^{(i_1'i_2'i_3')}}^{(3)}\big),
			\end{aligned}
		\end{multline*}
		whenever $j_{i_2}$ and $j_{i_2'}$ are both nonzero. When either of the two indices is null, the previous inequality is no longer true. In fact, if $j_{i_2}$ and $j_{i_2'}$ are both zero, the right-hand side vanishes, whereas the left-hand side can be positive. It is however easy to see that when $j_{i_2'} = 0$, the previous inequality can be made true by replacing $P_{i_2'}$ by $1/\sqrt{N}$ in the right-hand side (recalling that $P_{i_1i_2'i_3} = P_{i_1}\otimes P_{i_2'}\otimes P_{i_3}$). The same can also be done with $P_{i_2}$ when $j_{i_2} = 0$. In the following, it will often be the case that inequalities are proved under conditions such as $j_{i_2},j_{i_2'} \geq 1$ and can easily be adapted to cover $j_{i_2} = 0$ or $j_{i_2'} = 0$. We do not mention this each time to avoid redundancy. Using that
		\begin{equation}
			\label{eq:sum_J_upper_bound}
			\sum_{\underline{J}}\Gamma_{\underline{J}^{(i_1i_2i_3)}}^{(3)} \leq \Gamma^{(3)},
		\end{equation}
		we may carry out the $\underline{J}$ sum in \eqref{eq:low_momenta_estimate_expo_W3_far_first} to obtain
		\begin{multline*}
			\sum_{\underline{J}}\big\vert \Tr\big(P_{i_1i_2i_3}W_3(1,2,3)P_{i_1'i_2'i_3'}\Gamma_{\underline{J}^{(i_1i_2i_3;i_1'i_2'i_3')}}^{(3)}\big)\big\vert\\
			\begin{aligned}[t]
				&\leq CN^{(i_2 - i_1)\varepsilon/2}N^{(i_3 - i_1 + 4)\varepsilon/2}N^{(i_2' - i_1')\varepsilon/2}N^{(i_3' - i_1')\varepsilon/2}\Tr\big(P_{i_1i_2'i_3}(C + h_1)P_{i_1i_2'i_3}\Gamma^{(3)}\big)\\
				&\phantom{\leq} + CN^{(i_2 - i_1)\varepsilon/2}N^{(i_3 - i_1 + 4)\varepsilon/2}N^{(i_2' - i_1')\varepsilon/2}N^{(i_3' - i_1')\varepsilon/2}N^{-1}\Tr\big(P_{i_1i_3}(C + h_1)P_{i_1i_3}\Gamma^{(2)}\big)\\
				&\phantom{\leq} + CN^{(i_2 - i_1)\varepsilon/2}N^{(i_3 - i_1 + 4)\varepsilon/2}N^{(i_2' - i_1')\varepsilon/2}N^{(i_3' - i_1')\varepsilon/2}\Tr\big(P_{i_1'i_2i_3'}(C + h_1)P_{i_1'i_2i_3'}\Gamma^{(3)}\big)\\
				&\phantom{\leq} + CN^{(i_2 - i_1)\varepsilon/2}N^{(i_3 - i_1 + 4)\varepsilon/2}N^{(i_2' - i_1')\varepsilon/2}N^{(i_3' - i_1')\varepsilon/2}N^{-1}\Tr\big(P_{i_1'i_3'}(C + h_1)P_{i_1'i_3'}\Gamma^{(2)}\big).
			\end{aligned}
		\end{multline*}
		The two terms containing the factors $N^{-1}$ correspond to $j_{i_2} = 0$ and $j_{i_2'} = 0$. Lastly, carrying out the remaining sums using the geometric series formula, the resolution of the identity
		\begin{equation}
			\label{eq:resolution_identity_proof_low_occupancy}
			\sum_{i=1}^{M + 1}P_i = \mathds{1}
		\end{equation}
		and the fact that $M$ satisfies \eqref{eq:relation_M_epsilon}, we find
		\begin{equation}
			\label{eq:low_momenta_estimate_expo_W3_far_first_bound}
			\sum_{\underline{J}}\sum_{\substack{i_1 \geq i_2 \geq i_3\\ i_1 > i_3 + \Lambda}}\sum_{\substack{i_1' \geq i_2' \geq i_3'}}\big\vert\Tr\big(P_{i_1i_2i_3}W_3(1,2,3)P_{i_1'i_2'i_3'}\Gamma_{\underline{J}^{(i_1i_2i_3;i_1'i_2'i_3')}}^{(3)}\big)\big\vert \leq CN^{-(\Lambda - 3)\varepsilon/2}\Tr\big((C + h)\Gamma^{(1)}\big).
		\end{equation}
		Hence, we have bounded \eqref{eq:low_momenta_estimate_expo_W3_far_first} as desired.
		
		To bound \eqref{eq:low_momenta_estimate_expo_W3_far_second}, we begin by writing
		\begin{multline}
			\label{eq:low_momenta_estimate_expo_W3_far_second_intermediate}
			\big\vert \Tr\big(P_{i_1i_2i_3}W_3(1,2,3)P_{i_1'i_2'i_3'}\Gamma_{\underline{J}^{(i_1i_2i_3;i_1'i_2'i_3')}}^{(3)}\big)\big\vert \leq \tau CN^{2i_2\varepsilon}N^{-3}(j_{i_1} + 1)(j_{i_2} + 1)(j_{i_3} + 1)\Tr\Gamma_{\underline{J}^{(i_1i_2i_3)}}\\
			+ \tau^{-1}CN^{2i_1'\varepsilon}N^{-3}(j_{i_1'} + 1)(j_{i_2'} + 1)(j_{i_3'} + 1)\Tr\Gamma_{\underline{J}^{(i_1'i_2'i_3')}}.
		\end{multline}
		An appropriate choice of $\tau$ then gives
		\begin{multline}
			\label{eq:low_momenta_estimate_expo_W3_far_second_bound1}
			\big\vert \Tr\big(P_{i_1i_2i_3}W_3(1,2,3)P_{i_1'i_2'i_3'}\Gamma_{\underline{J}^{(i_1i_2i_3;i_1'i_2'i_3')}}^{(3)}\big)\big\vert \leq CN^{(i_1' - i_1 + 2)\varepsilon}\Tr\big(P_{i_1'i_2i_3}(C + h_2)P_{i_1'i_2i_3}\Gamma_{\underline{J}^{(i_1i_2i_3)}}^{(3)}\big)\\
			+ CN^{(i_1' - i_1 + 2)\varepsilon}\Tr\big(P_{i_1i_2'i_3'}(C + h_1)P_{i_1i_2'i_3'}\Gamma_{\underline{J}^{(i_1'i_2'i_3')}}^{(3)}\big),
		\end{multline}
		for $j_{i_1},j_{i_1'}\geq 1$. Although the adaptation for $j_{i_1'} = 0$ is straightforward - namely we just replace $P_{i_1'}$ by $1/\sqrt{N}$ - the case $j_{i_1} = 0$ is trickier. Indeed, we cannot simply replace $P_{i_1}$ by $1/\sqrt{N}$ because we also used the estimate $N^{2i_1\varepsilon}P_{i_1} \leq CP_{i_1}(C + h)P_{i_1}$. However, a different choice of $\tau$ in \eqref{eq:low_momenta_estimate_expo_W3_far_second_intermediate} yields
		\begin{multline}
			\label{eq:low_momenta_estimate_expo_W3_far_second_bound2}
			\big\vert \Tr\big(P_{i_1i_2i_3}W_3(1,2,3)P_{i_1'i_2'i_3'}\Gamma_{\underline{J}^{(i_1i_2i_3;i_1'i_2'i_3')}}^{(3)}\big)\big\vert \leq CN^{-(1 - \kappa)/2 + 2\varepsilon}\Tr\big(P_{i_2'i_2i_3}(C + h_2)P_{i_2'i_2i_3}\Gamma_{\underline{J}^{(i_1i_2i_3)}}^{(3)}\big)\\
			+ CN^{-(1 + \kappa)/2 + 2\varepsilon}\Tr\big(P_{i_1'i_3'}(C + h_1)P_{i_1'i_3'}\Gamma_{\underline{J}^{(i_1'i_2'i_3')}}^{(2)}\big),
		\end{multline}
		for $j_{i_1} = 0$. Combining \eqref{eq:low_momenta_estimate_expo_W3_far_second_bound1} and \eqref{eq:low_momenta_estimate_expo_W3_far_second_bound2}, we can carry out the sums in \eqref{eq:low_momenta_estimate_expo_W3_far_second} essentially as we did in \eqref{eq:low_momenta_estimate_expo_W3_far_first_bound} to obtain
		\begin{multline*}
			\sum_{\underline{J}}\sum_{\substack{i_1 \geq i_2 \geq i_3\\ \vert i_1 - i_3\vert \leq \Lambda}}\sum_{\substack{i_1' \geq i_2' \geq i_3'\\ \vert i_1' - i_3'\vert \leq \Lambda\\ i_1 > i_1' + \Lambda}}\big\vert\Tr\big(P_{i_1i_2i_3}W_3(1,2,3)P_{i_1'i_2'i_3'}\Gamma_{\underline{J}^{(i_1i_2i_3;i_1'i_2'i_3')}}^{(3)}\big)\big\vert\\
			\leq C\big(N^{-(\Lambda - 3)\varepsilon/2} + N^{-(1 - \kappa)/2 + 2\varepsilon}\big)\Tr\big((C + h)\Gamma^{(1)}\big).
		\end{multline*}
		This is the desired bound on \eqref{eq:low_momenta_estimate_expo_W3_far_second}.
		
		Having bounded \eqref{eq:low_momenta_estimate_expo_W3_far_first} and \eqref{eq:low_momenta_estimate_expo_W3_far_second}, all that is left to do to prove \eqref{eq:low_momenta_estimate_W3} is to deal with
		\begin{equation}
			\label{eq:low_momenta_estimate_expo_W3_close}
			\sum_{\underline{J}}\sum_{\substack{i_1 \geq i_2\geq i_3\\ \vert i_1 - i_3\vert\leq \Lambda}}\sum_{\substack{i_1'\geq i_2'\geq i_3'\\ \vert i_1' - i_3'\vert \leq \Lambda\\ \vert i_1 - i_1'\vert \leq \Lambda}}\big\vert\Tr\big(P_{i_1i_2i_3}W_3P_{i_1'i_2'i_3'}\Gamma_{\underline{J}^{(i_1i_2i_3;i_1'i_2'i_3')}}^{(3)}\big)\big\vert.
		\end{equation}
		Here, we are summing over $i_1 \geq \imax + 1$, meaning that we always have $j_{i_1} \leq N^{1 - \delta\varepsilon}$; we do not write it explicitly to avoid cluttering the notation. Proceeding as before, using additionally $j_{i_1} \leq N^{1 - \delta\varepsilon}$, we can write
		\begin{multline*}
			\big\vert\Tr\big(P_{i_1i_2i_3}W_3P_{i_1'i_2'i_3'}\Gamma_{\underline{J}^{(i_1i_2i_3;i_1'i_2'i_3')}}^{(3)}\big)\big\vert \leq CN^{-(\delta/2 - 2)\varepsilon}\sum_{i_3}\Tr\big(P_{i_2i_3}(C + h_1)P_{i_2i_3}\Gamma_{\underline{J}^{(i_1i_2i_3)}}^{(2)}\big)\\
			+ CN^{-(\delta/2 - 2)\varepsilon}\sum_{i_2' \geq i_3'}\Tr\big(P_{i_1'i_2'i_3'}(C + h_1)P_{i_1'i_2'i_3'}\Gamma_{\underline{J}^{(i_1'i_2'i_3')}}^{(3)}\big).
		\end{multline*}
		Thus, we can carry out all the sums in \eqref{eq:low_momenta_estimate_expo_W3_close} to obtain the bound
		\begin{equation*}
			\sum_{\underline{J}}\sum_{\substack{i_1 \geq i_2\geq i_3\\ \vert i_1 - i_3\vert\leq \Lambda}}\sum_{\substack{i_1'\geq i_2'\geq i_3'\\ \vert i_1' - i_3'\vert \leq \Lambda\\ \vert i_1 - i_1'\vert \leq \Lambda}}\big\vert\Tr\big(P_{i_1i_2i_3}W_3P_{i_1'i_2'i_3'}\Gamma_{\underline{J}^{(i_1i_2i_3;i_1'i_2'i_3')}}^{(3)}\big)\big\vert \leq CN^{-(\delta/2 - 2)\varepsilon}\Tr\big((C + h)\Gamma^{(1)}\big).
		\end{equation*}
		Gathering the previous estimates yields \eqref{eq:low_momenta_estimate_W3}.
		\bigskip
		
		\noindent
		\textit{Analysis of $W_2$.} We now prove \eqref{eq:low_momenta_estimate_W2}. We only provide the details of the proof for 
		\begin{equation*}
			W_2(1,2) = \p_1\cdot w_R(x_1 - x_2) + w_R(x_1 - x_2)\cdot \p_1
		\end{equation*}
		since it is the most difficult term to treat and that the others are bounded analogously. For the same reason, we take $i_1 \geq i_2 \geq i_2'$ and $i_1 \geq i_1' \geq i_2'$. To prove \eqref{eq:low_momenta_estimate_W2}, we distinguish between $\vert i_1' - i_2\vert \leq \Lambda$ and $\vert i_1' - i_2\vert > \Lambda$, and similarly for $i_1,i_2$ and $i_1,i_1'$. Since we only need to use the hypothesis $j_{i_1} \leq N^{1 - \delta\varepsilon}$ to deal with the case where the three indices $i_1,i_2,i_1'$ are close, we first take care of the others. Namely, we bound
		\begin{equation}
			\label{eq:low_momenta_estimate_expo_W2_far_first}
			\sum_{\underline{J}}\sum_{\substack{i_1 \geq i_2}}\sum_{\substack{i_1' \geq i_2'\\ i_2 \geq i_2'\\ i_1' > i_2 + \Lambda}}\sum_{i_3}\Tr\big(P_{i_1i_2i_3}W_2(1,2)P_{i_1'i_2'i_3}\Gamma_{\underline{J}^{(i_1i_2i_3;i_1'i_2'i_3)}}^{(3)}\big),
		\end{equation}
		\begin{equation}
			\label{eq:low_momenta_estimate_expo_W2_far_second}
			\sum_{\underline{J}}\sum_{\substack{i_1 > i_2 + \Lambda}}\sum_{\substack{i_1' \geq i_2'\\ i_2 \geq i_2'\\ \vert i_1' - i_2\vert \leq \Lambda}}\sum_{i_3}\Tr\big(P_{i_1i_2i_3}W_2(1,2)P_{i_1'i_2'i_3}\Gamma_{\underline{J}^{(i_1i_2i_3;i_1'i_2'i_3)}}^{(3)}\big)
		\end{equation}
		and
		\begin{equation}
			\label{eq:low_momenta_estimate_expo_W2_far_third}
			\sum_{\underline{J}}\sum_{\substack{i_1 \geq i_2\\ \vert i_1 - i_2\vert \leq \Lambda}}\sum_{\substack{i_1' \geq i_2'\\ i_2 \geq i_2'\\ \vert i_1' - i_2\vert \leq \Lambda\\ i_1 > i_1' + \Lambda}}\sum_{i_3}\Tr\big(P_{i_1i_2i_3}W_2(1,2)P_{i_1'i_2'i_3}\Gamma_{\underline{J}^{(i_1i_2i_3;i_1'i_2'i_3)}}^{(3)}\big).
		\end{equation}
		Note that there is only a single sum over the third coefficient $i_3$ (rather than a double sum) because of the orthogonality of the projections $P_i$.
		
		To bound \eqref{eq:low_momenta_estimate_expo_W2_far_first}, we start by writing
		\begin{multline*}
			\sum_{\substack{i_1,i_2'\\ i_2 \geq i_2'}}\Tr\big(P_{i_1i_2i_3}W_2(1,2)P_{i_1'i_2'i_3}\Gamma_{\underline{J}^{(i_1i_2i_3;i_1'i_2'i_3)}}^{(3)}\big)\\
			= \bigg\langle\sum_{i_1}P_{i_1i_2i_3}\otimes\mathds{1}^{\otimes (N-3)}\Psi_{\underline{J}^{(i_1i_2i_3)}},W_2(1,2)\sum_{\substack{i_2'\\ i_2 \geq i_2'}}P_{i_1'i_2'i_3}\otimes\mathds{1}^{\otimes(N-3)}\Psi_{\underline{J}^{(i_1'i_2'i_3)}}\bigg\rangle.
		\end{multline*}
		Then, thanks to
		\begin{equation*}
			\sum_{\substack{i_2'\\ i_2 \geq i_2'}}P_{i_1'i_2'i_3}\otimes\mathds{1}^{\otimes (N-3)}\Psi_{\underline{J}^{(i_1'i_2'i_3)}} = \bbP_M\otimes\bbP_{i_2}\otimes\mathds{1}^{\otimes(N-2)}\sum_{\substack{i_2'\\ i_2 \geq i_2'}}P_{i_1'i_2'i_3}\otimes\mathds{1}^{\otimes(N-3)}\Psi_{\underline{J}^{(i_1'i_2'i_3)}},
		\end{equation*}
		we can apply Lemma~\ref{lemma:plane_wave_estimates_W2_W3} and the Cauchy--Schwarz inequality to obtain
		\begin{multline*}
			\Big\vert\sum_{\substack{i_1,i_2'\\ i_2 \geq i_2'}}\Tr\big(P_{i_1i_2i_3}W_2(1,2)P_{i_1'i_2'i_3}\Gamma_{\underline{J}^{(i_1i_2i_3;i_1'i_2'i_3)}}^{(3)}\big)\Big\vert
			\leq \tau C\sum_{i_1}\Tr\big(P_{i_1i_2i_3}(-\Delta_1) P_{i_1i_2i_3} \Gamma_{\underline{J}^{(i_1i_2i_3)}}^{(3)}\big)\\
			+ \tau^{-1}CN^{2i_2\varepsilon}\sum_{\substack{i_2'\\ i_2 \geq i_2'}}\Tr\big(P_{i_1'i_2'i_3}\Gamma_{\underline{J}^{(i_1'i_2'i_3)}}^{(3)}\big),
		\end{multline*}
		for all $\tau > 0$. Taking $\tau = N^{(i_2 - i_1' + 1)\varepsilon}$ and using the estimate \eqref{eq:kinetic_part_lower_bound_N_2}, we deduce
		\begin{multline*}
			\Big\vert\sum_{\substack{i_1,i_2'\\ i_2 \geq i_2'}}\Tr\big(P_{i_1i_2i_3}W_2(1,2)P_{i_1'i_2'i_3}\Gamma_{\underline{J}^{(i_1i_2i_3;i_1'i_2'i_3)}}^{(3)}\big)\Big\vert \leq  CN^{(i_2 - i_1' + 1)\varepsilon}\sum_{i_1}\Tr\big(P_{i_1i_2i_3}(C + h_1)P_{i_1i_2i_3} \Gamma_{\underline{J}^{(i_1i_2i_3)}}^{(3)}\big)\\
			+ CN^{(i_2 - i_1' + 1)\varepsilon}\sum_{\substack{i_2'\\ i_2 \geq i_2'}}\Tr\big(P_{i_1'i_2'i_3}(C + h_1)P_{i_1'i_2'i_3}\Gamma_{\underline{J}^{(i_1'i_2'i_3)}}^{(3)}\big).
		\end{multline*}
		We may now carry out all the sums in \eqref{eq:low_momenta_estimate_expo_W2_far_first} 
		using \eqref{eq:sum_J_upper_bound} first, and then \eqref{eq:resolution_identity_proof_low_occupancy}, and the geometric series formula to obtain
		\begin{equation}
			\label{eq:low_momenta_estimate_expo_W2_far_first_bound}
			\Big\vert \sum_{\underline{J}}\sum_{\substack{i_1 \geq i_2}}\sum_{\substack{i_1' \geq i_2'\\ i_2 \geq i_2'\\ i_1' > i_2 + \Lambda}}\sum_{i_3}\Tr\big(P_{i_1i_2i_3}W_2(1,2)P_{i_1'i_2'i_3}\Gamma_{\underline{J}^{(i_1i_2i_3;i_1'i_2'i_3)}}^{(3)}\big)\Big\vert \leq N^{-\Lambda\varepsilon}\Tr\big((C + h)\Gamma^{(1)}\big).
		\end{equation}
		Thus, we have the bounded \eqref{eq:low_momenta_estimate_expo_W2_far_first} as wished.
		
		To bound \eqref{eq:low_momenta_estimate_expo_W2_far_second}, we this time write
		\begin{multline*}
			\Big\vert\sum_{\substack{i_2'\\ i_2 \geq i_2'}}\Tr\left(P_{i_1i_2i_3}W_2(1,2)P_{i_1'i_2'i_3}\Gamma_{\underline{J}^{(i_1i_2i_3;i_1'i_2'i_3)}}^{(3)}\right)\Big\vert \leq \tau CN^{2i_2\varepsilon}\Tr\big(P_{i_1i_2i_3} \Gamma_{\underline{J}^{(i_1i_2i_3)}}^{(3)}\big)\\
			+ \tau^{-1}C\sum_{\substack{i_2'\\ i_2 \geq i_2'}}\Tr\big(P_{i_1'i_2'i_3}(\p_1^\A)^2P_{i_1'i_2'i_3}\Gamma_{\underline{J}^{(i_1'i_2'i_3)}}^{(3)}\big),
		\end{multline*}
		which, by taking $\tau = N^{(i_1 - i_2 + 1)\varepsilon}$, further implies
		\begin{multline*}
			\Big\vert\sum_{\substack{i_2'\\ i_2 \geq i_2'}}\Tr\big(P_{i_1i_2i_3}W_2(1,2)P_{i_1'i_2'i_3}\Gamma_{\underline{J}^{(i_1i_2i_3;i_1'i_2'i_3)}}^{(3)}\big)\Big\vert \leq CN^{(i_2 - i_1 + 1)\varepsilon}\Tr\big(P_{i_1i_2i_3}(C + h_1)P_{i_1i_2i_3} \Gamma_{\underline{J}^{(i_1i_2i_3)}}^{(3)}\big)\\
			+ CN^{(i_2 - i_1 + 1)\varepsilon}\sum_{\substack{i_2'\\ i_2 \geq i_2'}}\Tr\big(P_{i_1'i_2'i_3}(C + h_1)P_{i_1'i_2'i_3}\Gamma_{\underline{J}^{(i_1'i_2'i_3)}}^{(3)}\big).
		\end{multline*}
		Carrying out all the remaining sums as in \eqref{eq:low_momenta_estimate_expo_W2_far_first_bound}, we obtain
		\begin{equation*}
			\Big\vert \sum_{\underline{J}}\sum_{\substack{i_1 > i_2 + \Lambda}}\sum_{\substack{i_1' \geq i_2'\\ i_2 \geq i_2'\\ \vert i_1' - i_2\vert \leq \Lambda}}\sum_{i_3}\Tr\big(P_{i_1i_2i_3}W_2(1,2)P_{i_1'i_2'i_3}\Gamma_{\underline{J}^{(i_1i_2i_3;i_1'i_2'i_3)}}^{(3)}\big)\Big\vert \leq N^{-\Lambda\varepsilon}\Tr\big((C + h)\Gamma^{(1)}\big),
		\end{equation*}
		which is the desired estimate on \eqref{eq:low_momenta_estimate_expo_W2_far_second}.
		
		To bound \eqref{eq:low_momenta_estimate_expo_W2_far_third}, we use Lemma~\ref{lemma:state_three_projections_full_trace} to obtain
		\begin{multline}
			\label{eq:low_momenta_estimate_expo_W2_far_third_bound_intermediate}
			\Big\vert\sum_{\substack{i_2'\\ i_2 \geq i_2'}}\Tr\big(P_{i_1i_2i_3}W_2(1,2)P_{i_1'i_2'i_3}\Gamma_{\underline{J}^{(i_1i_2i_3;i_1'i_2'i_3)}}^{(3)}\big)\Big\vert \leq \tau CN^{2i_2\varepsilon}N^{-3}(j_{i_1} + 1)(j_{i_2} + 1)(j_{i_3} + 1)\Tr\Gamma_{\underline{J}^{(i_1i_2i_3)}}\\
			+ \tau^{-1}CN^{2i_1'\varepsilon}\sum_{\substack{i_2'\\ i_2 \geq i_2'}}N^{-3}(j_{i_1'} + 1)(j_{i_2'} + 1)(j_{i_3} + 1)\Tr\Gamma_{\underline{J}^{(i_1'i_2'i_3)}},
		\end{multline}
		for all $\tau > 0$. Choosing $\tau$ such that $j_{i_1}$ and $j_{i_1'}$ are swapped and balancing out the front coefficients, we find
		\begin{multline*}
			\Big\vert\sum_{\substack{i_2'\\ i_2 \geq i_2'}}\Tr\big(P_{i_1i_2i_3}W_2(1,2)P_{i_1'i_2'i_3}\Gamma_{\underline{J}^{(i_1i_2i_3;i_1'i_2'i_3)}}^{(3)}\big)\Big\vert \leq CN^{(i_1' - i_1 + 2)\varepsilon}\Tr\big(P_{i_1'i_2i_3}(C + h_2)P_{i_1'i_2i_3}\Gamma_{\underline{J}^{(i_1i_2i_3)}}^{(3)}\big)\\
			+ CN^{(i_1' - i_1 + 2)\varepsilon}\Tr\big(P_{i_1i_2'i_3}(C + h_1)P_{i_1i_2'i_3}\Gamma_{\underline{J}^{(i_1'i_2'i_3)}}^{(3)}\big).
		\end{multline*}
		Here, the adaptation for $j_{i_1'} = 0$ is straightforward, whereas it is more delicate for $j_{i_1} = 0$. More specifically, when $j_{i_1} = 0,$ we instead write
		\begin{multline*}
			\Big\vert\sum_{\substack{i_2'\\ i_2 \geq i_2'}}\Tr\big(P_{i_1i_2i_3}W_2(1,2)P_{i_1'i_2'i_3}\Gamma_{\underline{J}^{(i_1i_2i_3;i_1'i_2'i_3)}}^{(3)}\big)\Big\vert \leq CN^{-(1 - \kappa)/2 + 2\varepsilon}\Tr\big(P_{i_2i_3}(C + h_1)P_{i_2i_3}\Gamma_{\underline{J}^{(i_1i_2i_3)}}^{(2)}\big)\\
			+ CN^{-(1 + \kappa)/2 + 2\varepsilon}\sum_{\substack{i_2'\\ i_2 \geq i_2'}}\Tr\big(P_{i_1'i_2'i_3}(C + h_1)P_{i_1'i_2'i_3}\Gamma_{\underline{J}^{(i_1'i_2'i_3)}}^{(2)}\big).
		\end{multline*}
		Carrying out the sums in \eqref{eq:low_momenta_estimate_expo_W2_far_third}, we obtain
		\begin{multline*}
			\Big\vert \sum_{\underline{J}}\sum_{\substack{i_1 \geq i_2\\ \vert i_1 - i_2\vert \leq \Lambda}}\sum_{\substack{i_1' \geq i_2'\\ i_2 \geq i_2'\\ \vert i_1' - i_2\vert \leq \Lambda\\ i_1 > i_1' + \Lambda}}\sum_{i_3}\Tr\big(P_{i_1i_2i_3}W_2(1,2)P_{i_1'i_2'i_3}\Gamma_{\underline{J}^{(i_1i_2i_3;i_1'i_2'i_3)}}^{(3)}\big)\Big\vert\\
			\leq C\big(N^{-(\Lambda - 3)\varepsilon/2} + N^{-(1 - \kappa)/2 + 2\varepsilon}\big)\Tr\big((C + h)\Gamma^{(1)}\big).
		\end{multline*}
		
		Having bounded \eqref{eq:low_momenta_estimate_expo_W2_far_first}--\eqref{eq:low_momenta_estimate_expo_W2_far_third}, we are left with
		\begin{equation}
			\label{eq:low_momenta_estimate_expo_W2_close}
			\sum_{\underline{J}}\sum_{\substack{i_1 \geq i_2\\ \vert i_1 - i_2\vert \leq \Lambda}}\sum_{\substack{i_1' \geq i_2'\\ i_2 \geq i_2'\\ \vert i_1' - i_2\vert \leq \Lambda\\ \vert i_1' - i_1 \vert \leq \Lambda}}\sum_{i_3}\Tr\big(P_{i_1i_2i_3}W_2(1,2)P_{i_1'i_2'i_3}\Gamma_{\underline{J}^{(i_1i_2i_3;i_1'i_2'i_3)}}^{(3)}\big).
		\end{equation}
		Using \eqref{eq:low_momenta_estimate_expo_W2_far_third_bound_intermediate}, $j_{i_1} \leq N^{1 - \delta\varepsilon}$ and balancing out the front coefficients, we find
		\begin{multline*}
			\Big\vert\sum_{\substack{i_2'\\ i_2 \geq i_2'}}\Tr\big(P_{i_1i_2i_3}W_2(1,2)P_{i_1'i_2'i_3}\Gamma_{\underline{J}^{(i_1i_2i_3;i_1'i_2'i_3)}}^{(3)}\big)\Big\vert \leq CN^{-\delta\varepsilon/2 + \varepsilon}\Tr\big(P_{i_2i_3}(C + h_1)P_{i_2i_3}\Gamma_{\underline{J}^{(i_1i_2i_3)}}^{(2}\big)\\
			+ CN^{-\delta\varepsilon/2 + \varepsilon}\sum_{\substack{i_2'\\ i_2 \geq i_2'}}\Tr\big(P_{i_1'i_2'i_3}(C + h_1)P_{i_1'i_2'i_3}\Gamma_{\underline{J}^{(i_1'i_2'i_3)}}^{(2}\big).
		\end{multline*}
		Finally, we obtain
		\begin{equation*}
			\Big\vert \sum_{\underline{J}}\sum_{\substack{i_1 \geq i_2\\ \vert i_1 - i_2\vert \leq \Lambda}}\sum_{\substack{i_1' \geq i_2'\\ i_2 \geq i_2'\\ \vert i_1' - i_2\vert \leq \Lambda\\ \vert i_1' - i_1 \vert \leq \Lambda}}\sum_{i_3}\Tr\big(P_{i_1i_2i_3}W_2(1,2)P_{i_1'i_2'i_3}\Gamma_{\underline{J}^{(i_1i_2i_3;i_1'i_2'i_3)}}^{(3)}\big)\Big\vert \leq CN^{-(\delta/2 - 2)\varepsilon}\Tr\big((C + h)\Gamma^{(1)}\big).
		\end{equation*}
		
		Gathering the previous estimates yields \eqref{eq:low_momenta_estimate_W2}, thereby concluding the proof of Lemma~\ref{lemma:low_occupancy_estimates}.
	\end{proof}
	
	\printbibliography
	
\end{document}